\documentclass[fleqn,usenatbib]{mnras}

\usepackage{mathptmx}

\usepackage[T1]{fontenc}
\usepackage{ae,aecompl}

\DeclareRobustCommand{\VAN}[3]{#2}
\let\VANthebibliography\thebibliography
\def\thebibliography{\DeclareRobustCommand{\VAN}[3]{##3}\VANthebibliography}

\usepackage{graphicx}	
\usepackage{amsmath}	
\usepackage{amssymb}	

\usepackage[normalem]{ulem}
\usepackage{booktabs}

\usepackage{lmodern}

\newcommand{\rdos}{$R_{200}$}

\newcommand\sag {\textsc{sag}}
\newcommand\mdpl {\textsc{mdpl2}}

\title[SF quenching in the \textsc{The Three Hundred}]{The Three Hundred project: connection between star formation quenching and dynamical evolution in and around simulated galaxy clusters}

\author[T. Hough et al.]{
Tomás Hough,$^{1,2}$\thanks{E-mail: tomashough@gmail.com}
Sofía A. Cora,$^{1,2}$
Roan Haggar,$^{3,4}$
Cristian Vega-Martinez,$^{5,6}$
Ulrike Kuchner,$^{3}$
\newauthor{
Frazer Pearce,$^{3}$
Meghan Gray,$^{3}$
Alexander Knebe,$^{7,8,9}$
and Gustavo Yepes$^{8,9}$
}
\\
$^{1}$Instituto de Astrof\'isica de La Plata (CCT La Plata, CONICET, UNLP), Paseo del Bosque s/n, La Plata, Argentina\\
$^{2}$Facultad de Ciencias Astron\'omicas y Geof\'isicas, Universidad Nacional de La Plata, Paseo del Bosque s/n, La Plata, Argentina\\
$^{3}$School of Physics \& Astronomy, University of Nottingham, Nottingham NG7 2RD, UK\\
$^{4}$Waterloo Centre for Astrophysics, University of Waterloo, Waterloo, Ontario N2L 3G1, Canada\\
$^{5}$ Instituto de Investigaci\'on Multidisciplinar en Ciencia y Tecnolog\'ia, Universidad de La Serena, Ra\'ul Bitr\'an 1305, La Serena, Chile\\
$^{6}$ Departamento de Astronom\'ia, Universidad de La Serena, Av. Juan Cisternas 1200 Norte, La Serena, Chile\\
$^{7}$ Departamento de Física Teórica, Módulo 8, Facultad de Ciencias, Universidad Autónoma de Madrid, 28049 Madrid, Spain\\
$^{8}$ Centro de Investigación Avanzada en Física Fundamental (CIAFF), Facultad de Ciencias, Universidad Autónoma de Madrid, Spain\\
$^{9}$ International Centre for Radio Astronomy Research, University of Western Australia, 35 Stirling Highway, Crawley, Australia
}

\date{Accepted XXX. Received YYY; in original form ZZZ}

\pubyear{2022}

\begin{document}
\label{firstpage}
\pagerange{\pageref{firstpage}--\pageref{lastpage}}
\maketitle

\begin{abstract}
In this work, we combine the semi-analytic model of galaxy formation and evolution \sag~with the $102$ relaxed simulated galaxy clusters 
from \textsc{The Three Hundred} project, and we study the link between the quenching of star formation (SF) and the physical processes that galaxies experience through their dynamical history in and around clusters. 
We classify galaxies in four populations based on their orbital history: recent and ancient infallers, and
backsplash and neighbouring galaxies. 
We find that $\sim 85$ per cent of the current population of quenched galaxies located inside the clusters are ancient infallers with low or null content of hot and cold gas. The fraction of quenched ancient infallers increases strongly between the first and second pericentric passage, due to the removal of hot gas by the action of ram-pressure stripping (RPS). 
The majority of them quenches after the first pericentric passage, but a non-negligible fraction needs a second passage, specially galaxies with $M_\star \leq 10^{10.5} \, {\rm M_\odot}$.
Recent infallers represent $\sim 15$ per cent of the quenched galaxies located inside the cluster and, on average, they contain a high proportion of hot and cold gas; moreover, pre-processing effects are the responsible for quenching the recent infallers prior to infall onto the main cluster progenitor. 
The $\sim 65$ per cent of quenched galaxies located around clusters are backsplash galaxies, for which the combination of RPS acting during a pre-processing stage and inside the cluster is necessary for the suppression of SF in this population.

\end{abstract}

\begin{keywords}
galaxies: clusters: general -- galaxies: evolution -- methods: numerical
\end{keywords}



\section{Introduction}
\label{sec:intro}

The star formation (SF) activity of a galaxy depends 
on both its stellar mass content and the local density of the environment where it resides \citep{peng2010}. 
Galaxies located in high-density environments are more likely to be quiescent, and to have redder colours and more spheroidal morphology than galaxies with similar stellar mass located in the field, both at high redshift \citep{muzzin2013,tomczak2016,fossati2017,sherman2020} and in the local Universe \citep{Wetzel12,bluck2016, jian2017, davies2019}. 
In addition, high-mass galaxies tend to have these characteristics as well, regardless of the environment in which they reside \citep[][]{kauffmann2004,weinmann2006,Wetzel12,Wetzel13}.

Once a galaxy enters a high-density environment like a cluster, physical processes associated to the environment such as ram-pressure stripping \citep[RPS,][]{gg72,abadi1999}, tidal stripping \citep[TS,][]{merritt1983}, galaxy harassment \citep{moore1996} or starvation \citep{larson1980,balogh2000} start to (mainly) affect the gas content of the galaxy, hence altering its properties. From this, 
the ram-pressure (RP) exerted by the intracluster medium (ICM)
over the gaseous components of the galaxy
has been proposed as the main driver of galaxy transformation once the galaxy becomes a satellite \citep{mccarthy2008,tecce10, bahe2013,jaffe15,roberts19}, by removing the hot gas from the halo and/or the cold gas from the disc, and hence 
reducing or quenching its SF.
Evidence of extreme RPS acting over the cold gas phase of disc galaxies \citep[which includes the visually striking `jellyfish' galaxies,][]{ebeling2014} has been found to occur in massive clusters both observationally
\citep{chung2009,poggianti2017,boselli2018c,roberts2020,Moretti_2021} and theoretically \citep{yun2019,Franchetto_2021}. We refer the reader to \citet{boselli2021} for a recent review on the topic.
RP depends on both the local ICM density and relative velocity of the galaxy, for which it is expected to be more intense in high-mass clusters and over galaxies moving at high relative velocities. However, the timescales involved in the action of 
RP are not very well constrained. While some authors find that this timescale is short, of 
the order of $\sim 1\,{\rm Gyr}$ or less
\citep{mccarthy2008, catinella2013, Lotz_2019}, others find that the whole process of quenching a galaxy takes several Gyr \citep{delucia2012,Wetzel13,Oman16,cora2019}. Thus, local environmental density results insufficient to determine galaxy properties, as they seem to depend also on the time a galaxy spend inside that environment \citep{smith2019}.

The phase-space diagram (PSD), constructed with the positions and velocities of galaxies relative to the cluster centre, has been proven to be successful in constraining the accretion epochs and orbital histories of different cluster galaxy populations \citep{oman13, smith2019, delosrios21} and to link
environment and galaxy properties \citep{mahajan11,haines15, pasquali2019, rhee2020,sampaio2021}.
In several of the aforementioned studies, the combination of observed
PSD and numerical simulations has been used to identify different populations of galaxies in and around galaxy clusters, such as an infalling population (galaxies with high velocities that are approaching the cluster for the first time), a virialized population (galaxies that have entered the cluster several Gyr ago, have lower relative velocities and are near the cluster centre) and a backsplash population \citep[galaxies that have been inside the cluster and are currently located outside of it,][]{balogh2000, gill05}. 
However, the limitations due to projection in observed PSD redounds in uncertainties in the classification of galaxy populations. For instance, backsplash galaxies
and infallers occupy regions that overlap in the projected PSD \citep{mahajan11, oman13, delosrios21}, and interlopers can account for a non-negligible proportion of galaxies throughout the radial distribution \citep{owers2019,delosrios21}. 

Massive clusters ($M_{\rm 200}>10^{14.5} \, {\rm M_\odot}$)
\footnote{The mass $M_{200}$ is estimated by considering an homogeneous mass distribution enclosed by a characteristic radius of a dark matter halo, $R_{200}$, at which the density is equal to $200$ times the critical density of the Universe at the redshift of the system. When the mean density of the Universe is used instead, we refer to this radius as $R_{\rm 200m}$.}
accrete galaxies that inhabit groups ($M_{\rm 200}\sim 10^{13} \, {\rm M_\odot}$) or filamentary
environments. Using a semi-analytic approach, \citet{mcgee2009} show that a massive cluster accretes $\sim 40$ per cent of its galaxies from group-like environments, while \citet{benavides2020} use the Illustris hydrodynamical simulation \citep{vogelsberger2014} and find that $\sim 38$ per cent of $z=0$ satellites were accreted as part of groups, although these fractions are strongly dependent on the definition of `group'. Furthermore, \citet{Wetzel13} find that, in the most massive host haloes,
the dominant mode of infall is as satellite in lower mass haloes, as the fraction of centrals accreted from the field drops below $\sim 50$ per cent. 
Physical processes that satellites experience prior to infall are considered as `pre-processing' \citep{balogh1999, Wetzel13, smith2019,pallero2019,davies2019}. 
As an example, \citet{fabello2012} found evidence that RP strips atomic gas of galaxies in haloes more massive than $10^{13} {\rm M_\odot}$, whereas this should not be expected in lower mass haloes \citep{taranu2014}. In addition, \citet{dzudzar2021} suggest that tidal interactions are the primary cause of the 
H\textsc{\,i}
deficiency in group galaxies, while \citet{oh2018} show evidence that mergers prior to infall can induce morphological transformation in recently accreted cluster galaxies.
By means of the Yosei Zoom-in Cluster Simulations (YZiCS), which consist in the hydrodynamical simulations of 16 galaxy clusters that span masses of $5\times 10^{13}<M_{200}/{\rm M_\odot}<10^{15}$, \citet{jung2018} study the excess of gas-poor galaxies in cluster environments. By investigating the radial trends of cold gas content, they find that the fraction of gas-depleted galaxies increases with decreasing distance inside $1.5 \, R_{200}$. Moreover, they find evidence of pre-processing, as $\sim 30$ per cent of cluster galaxies 
are already gas-poor before being accreted into the cluster,
and $\sim 45$ per cent of those galaxies infalling as satellites have a similar condition.
\cite{Wetzel12} found that $\sim 40-55$ per cent of galaxies in the outskirts of clusters are quenched, so a precise classification of the galaxy population around galaxy clusters
is fundamental to disentangle between backsplash galaxies
that have been quenched by the cluster, and infallers that could
have been quenched as a result of pre-processing. 

Several observational studies have shown that environmental effects might act beyond the halo boundary over baryonic components \citep{Wetzel12, lu2012} and 
dark matter (DM)
haloes \citep{behroozi2014}. 
From a sample of simulated galaxy clusters constructed 
from full-physics hydrodynamical re-simulations of the sample of clusters of 
\textsc{The Three Hundred} Project
by using the Smooth-Particle-Hydrodynamics code \textsc{Gadget-X}
\citep{Cui_2018} and the use of the PSD,
\citet{arthur19} 
show that first infallers lose their gas at $\sim 1.5 \, R_{200}$
in real space, whereas the cut-off in projected coordinates occurs closer to $2 \, R_{200}$. \citet{mostoghiu21} extend this study and explore how galaxies accreted at different epochs
build-up the PSD, finding that galaxies located within $\sim 1.5 \, R_{200}$ have lost nearly all of their gas, regardless of their time since infall into the cluster. On the contrary, \citet{sun2007} and \citet{jeltema2008} find evidence of satellite galaxies located in the central regions of clusters that present stripping of cold gas and of the hot gas halo (respectively), showing that not all satellites are completely stripped of their halo gas.  
Recently, \citet{ayromlou2019} introduce a new technique to estimate environmental effects in the Munich semi-analytic model \textsc{L-galaxies}. They define a local background density estimator associated to the environment where galaxies inhabit, regardless whether
they are central or satellites. This devises a new RPS model, in which central galaxies located in the vicinity of massive haloes ($M_{200}>10^{12} \, {\rm M_\odot}$) can experience RPS of their hot gas, and show that the dependence of the fraction of quenched galaxies on
cluster-centric distance up to $6 \, R_{200}$ is in reasonable agreement with observations \citep{Ayromlou_2021b}. However, \citet{Wetzel_2014} argue that the evolution of the 
star formation rate (SFR)
and quenching of backsplash galaxies occur similarly as in satellites that remain within their host halo, which can explain the enhanced quiescent fraction of central galaxies near massive host haloes up to $\sim 5 \, R_{\rm 200m}$.
With the results of their model, 
SF quenching produced by the environment in which galaxies reside (environmental quenching) is restricted to the region within the virial radius of the host halo, and no additional environmental effects can act over central galaxies prior to the first infall into the cluster.

The infall time into galaxy clusters
results a crucial parameter to study galaxy evolution in these environments. To give an insight on the satellites that orbit inside the virial radius of their host, \citet{smith2019} study the physical properties of galaxies from Sloan Digital Sky Survey \citep[SDSS, ][]{abazajian2009} in the projected PSD. They classify satellites according to their lookback
time since first infall into the host halo, $t_{\rm lbk,infall}$
and define a `recent infaller' sample and an `ancient infaller' one, with 
$t_{\rm lbk,infall}<3\,{\rm Gyr}$ and $t_{\rm lbk,infall}>5\,{\rm Gyr}$, respectively. 
They find that ancient infallers are, on average, characterized by small pericentric distances, 
i.e. fractions of the virial radius
$r_{\rm per}/R_{\rm vir} \sim 0.2-0.3$, 
while recent infallers have $r_{\rm per}/R_{\rm vir} \sim 0.6$. In addition, ancient infallers typically reach $\sim 90$  per cent of their final stellar mass $\sim 2\,{\rm Gyr}$ earlier than recent infallers do, meaning that galaxies that infall into hosts earlier are quenched earlier. Moreover, recent infallers seem
to be affected by the large scale environment prior to infall, as they display reduced stellar mass growth $\sim 9\,{\rm Gyr}$ ago. Using a similar approach, \citet{pasquali2019} find that the global specific star formation rate (sSFR; defined as the ratio between 
the SFR and stellar mass)
of low-mass satellites declines more rapidly than that of high-mass satellites. 
Hence, it is key to determine the time of infall of galaxies into high-density environments precisely
for the correct interpretation of the SF quenching and the identification of the physical processes involved.
Recently, \citet{delosrios21} developed a machine learning algorithm trained to recover a 3D orbital classification of galaxies based on the position on the projected PSD. This novel technique allows to disentangle between backsplash galaxies,
interlopers, and satellites with different infall times.

In this work, we combine the DM-only simulations of galaxy clusters of 
\textsc{The Three Hundred}
project with the semi-analytic model of galaxy formation \sag~ \citep[acronym for `Semi-Analytic Galaxies'][]{Cora_2018} to study the physical mechanisms involved in the SF quenching process 
of galaxies in and around clusters
and the connection with their dynamical evolution.
This work complements the research carried out in
\citet{cora2019}, 
which is focused on the relevance of environmental quenching and mass quenching (which refers to any internal process dependent on the galaxy stellar mass) in the suppression of SF, by using the \mdpl~DM-only MultiDark cosmological simulation. Thus, we limit the analysis to the \sag~data exclusively.

This work is organized as follows.
In Sec.~\ref{sec:simulation}, we provide details of the DM-only simulations and we describe those physical processes included in the \sag~model that are more relevant for this work. In Sec.~\ref{sec:populations}, we give a general description of the spatial distribution 
in and around the galaxy clusters of different galaxy populations, classified as centrals and satellites, on the one hand, and recent infallers, ancient infallers and backsplash galaxies, on the other.
In Sec.~\ref{sec:AINRINBS}, we analyze the SF activity, the hot gas content and the position in the PSD of the different galaxy populations classified by their orbital evolution, and determine the relevance of the first pericentric passage on the SF quenching process. In Sec.~\ref{sec:transformation}, we 
estimate the amount of hot gas removed by RPS
in different stages along the orbital evolution. In Sec.~\ref{sec:discussion}, we discuss our results in the light of similar works from the literature.  Sec.~\ref{sec:conclusions} contains a summary of our main conclusions.

\section{Sample of simulated galaxy clusters}
\label{sec:simulation}

Our sample of simulated galaxy clusters is constructed by combining DM-only simulations of regions containing 
cluster-like haloes
and the semi-analytic model of galaxy formation \sag.
The  DM simulations are taken from 
{\textsc The Three Hundred} project that offers a dataset which consists of $324$ regions containing the most massive clusters at $z=0$ present in the DM-only cosmological simulation \mdpl.
We focus our study on relaxed galaxy clusters, which can be considered as large cosmological laboratories to study how 
physical processes associated with high-density environments affect galaxy properties. 
In this section, we describe the simulation \mdpl, define the criteria to select haloes that will host relaxed clusters and present the main characteristics of the \sag~model.

\subsection{DM-only simulation \mdpl~and cluster-like regions}

The \mdpl~DM-only MultiDark simulation \citep{Klypin16} comprises $3840^3$ particles with mass of $1.5\times 10^9 \rm M_\odot$ distributed inside a box with sides of $1\,h^{-1}\, {\rm Gpc}$.
It is consistent with a flat $\Lambda$CDM model characterized by Planck cosmological parameters:
$\Omega_{\rm m}$~=~0.307, $\Omega_\Lambda$~=~0.693, 
$\Omega_{\rm B}$~=~0.048, $n_{\rm s}$~=~0.96
and $H_0$~=~100~$h^{-1}$~km~s$^{-1}$~Mpc$^{-1}$, where $h$~=~0.678 
\citep{Planck2016}.

The dataset of simulated galaxy clusters is built by extracting $324$ spherical regions from the \mdpl~simulation. Each of these regions is centred on the most massive DM haloes identified at $z=0$ by the $\textsc{roskstar}$ halo finder \citep{Behroozi_rockstar}
and has a radius of $15 \, {h}^{-1} \, {\rm Mpc}$, which represents $\sim 10\, R_{200}$. These massive DM haloes have masses 
$M_{200} > 10^{14.95}\, {\rm M}_\odot$ \citep{Cui_2018}.
They are main host haloes that host galaxy clusters and are surrounded by many additional objects like smaller haloes (that host galaxy groups) and filaments, which allow to study not only the population of galaxies of the clusters themselves but also the ones present in the environment where the clusters reside. 
The DM halo catalogues in \textsc{The Three Hundred} project were build with the Amiga Halo Finder \citep[AHF,][]{knollmannAHF2009} and,
for this work,
the halo catalogues and merger trees were re-constructed with \textsc{ConsistenTrees} \citep{Behroozi_ctrees} because of particular 
requirements
of the \textsc{sag} model.

\subsection{Sample of relaxed cluster-like haloes}

The complete sample of the $324$ spherical regions of \textsc{The Three Hundred} project 
provides a dynamically heterogeneous sample of cluster-like haloes (to which we simply refer to as `clusters' in the rest of this section).
\cite{Cui_2018} characterize their dynamical relaxation state based on three parameters:

\begin{itemize}
    \item The virial ratio $\eta = (2\,T-E_{\rm s})/|W|$, that measures how well a cluster obeys the virial theorem, where $T$ is the total kinetic energy, $E_{\rm s}$ is the energy from surface pressure and $W$ is its total potential energy.
    \item Centre of mass offset, $\Delta_{\rm r} =  |R_{\rm cm}-R_{\rm c}|/R_{200}$, where $ R_{\rm cm}$ is the centre of mass of a cluster of radius $R_{200}$, and $ R_{\rm c}$ is the maximum density peak of the halo.
    \item Subhalo mass fraction, $f_{\rm s}$, which is the fraction of the cluster mass that is contained in subhaloes.
\end{itemize}

\citet{Cui_2018} consider that a cluster is relaxed if it satisfies that $\Delta_{\rm r}<0.04$, $f_{\rm s}<0.1$ and $0.85<\eta<1.15$. To determine the relaxation state of each cluster, we choose to follow the approach established in \citet{haggar2020}, where they define a continuous parameter $\chi_{\rm DS}$ that combines the three parameters previously defined, i.e. 

\begin{equation}
    \chi_{\rm DS} = \sqrt{\frac{3}{\left(\frac{\Delta_{\rm r}}{0.04}\right)^2+\left(\frac{f_{\rm s}}{0.01}\right)^2 + \left(\frac{|1-\eta|}{0.15}\right)^2 } },
\end{equation}

\noindent that separates the most relaxed ($\chi_{\rm DS}>1.03$) from the least relaxed clusters ($\chi_{\rm DS}<0.619$) at $z=0$.
As our goal is to explore the link between the evolution of the 
SF activity of galaxies and their dynamical state, we choose to carry out our analysis based on the most relaxed clusters. 
The RPS model implemented in \sag~assumes 
spherical symmetric mass distribution for the host haloes,
and the impact on the galaxy population is likely to be more realistic in relaxed clusters since they are unlikely to have undergone a major merger recently \citep{contreras2022}, and tend to have more spherical morphology.
Hence, 
from the complete cluster sample, we extract those clusters that have $\chi_{\rm DS}>1$\footnote{We choose this value for simplicity; it does not affect the number of relaxed clusters with respect to that obtained by using $\chi_{\rm DS}>1.03$.}, so that our sample consists of $102$ relaxed clusters.
The sample comprises
halo masses ${\rm log}(M_{200}[{\rm M}_\odot])>14.95$.

\subsection{Semi-analytic model \sag}
\label{sec:model}

The semi-analytic model of galaxy formation and evolution \sag~includes usual physical processes: radiative cooling of hot gas, quiescent star formation and starbursts triggered by disc instabilities and galaxy mergers, chemical enrichment, feedback from supernova (SN) explosions, growth of supermassive black holes in galaxy centres and the consequent
feedback form active galactic nucleus (AGN).
It originates in the Munich model \citep{Springel_2001}, and has been highly modified as described in \citet{Cora_2006}, \citet{lcp08}, \citet{MunozArancibia_2015}, \citet{Gargiulo_2015}, and \citet{Cora_2018}.
The latter work describes the most recent version of the model, which
includes the action of 
RPS on the hot gas and cold gas components of satellite galaxies \citep{tecce10,tecce11,vega2022}, which is particularly relevant for this study.
This process require an adequate tracking of the satellite orbits. This is directly provided by the DM-only simulations for those satellites that keep their DM haloes. 
However, in many cases, 
DM substructures are not detected by the halo finder because of the limited numerical resolution of the simulation or substructures may be disrupted, although the origin of most disruptions of subhaloes identified in cosmological simulations is numerical \citep{vdBosch_Ogiya_2018}. Satellite galaxies hosted by those subhaloes that become unresolved are known as orphan galaxies,
and we integrate their orbits analitically in a pre-processing step, before applying \sag.
In the following subsections, we describe briefly the most relevant
physical processes to this work, i.e. the SF process, the environmental effects and the orbit integration of orphan galaxies; we refer the reader to \citet{Cora_2018} for a complete description of the rest of the physical processes included in the model and information about the values assigned to the free parameters that regulate the efficiency of the different processes.

The galaxy populations generated by applying \sag~to the DM-only simulations of the $102$ relaxed cluster-like DM haloes
are stacked and considered as a single sample in order to increase the statistics of our subsequent analysis.
In order to avoid resolution effects, we select galaxies with stellar mass ${\rm log}(M_\star \,{[\rm M_\odot]})\geq 9$, which results in halo masses of ${\rm log}(M_{200}\, {[\rm M_\odot]})\geq 10.64$, in consistency with the threshold imposed by \citet{haggar2020}.

\subsection*{Gas reservoirs}
\label{subsec:gasreservoirs}

Cosmic accretion of baryons fills main host DM haloes with gas that is shocked and heated. We compute the mass of this hot gas as

\begin{equation}
  \begin{split}
    M_\text{hot} = & f_{\rm b} M_\text{200} - M_{\star,\text{cen}} -
    M_\text{cold,cen} - M_\text{BH,cen} \\
    & - \sum_{i=1}^{N_\text{sat}} \left( M_{\star,i} +
    M_{\text{cold,}i} + M_{\text{hot,}i} + M_{\text{BH,}i} \right),
  \end{split}
  \label{eq:mhot}
\end{equation}

\noindent where $f_{\rm b}=0.1569$ is the baryon fraction, 
$M_{\rm 200}$
characterises the virial mass of the main host halo, and
$M_\star$, $M_{\rm hot}$, $M_{\rm cold}$ and $M_{\rm BH}$ represent the stellar mass, hot gas mass, cold gas mass and super massive black hole mass of the central or satellite galaxies of the host halo, respectively.  The hot gas is distributed following an isothermal sphere, and both centrals and satellites that still preserve a hot gas reservoir will proceed to cool gas. 
Gas cooling rates are calculated based on the model of \citet{Springel_2001}, considering the radiated power per chemical element given in \citet{foster2012}.
As the gas cools and falls into the potential well of the halo, it forms a pressure-supported disc of cold gas where new stars are formed.

\subsection*{SF scheme}
\label{subsec:SF}

SF takes place in both quiescent mode and starburst mode.
The quiescent SF model is based on the implementation done in \citet{croton2016}, first formulated in \citet{croton2006}. Motivated by the empirical/observational relation of \citet{kennicutt1998}, the model adopts a cold gas surface density threshold below which no gas is converted into stars. Assuming that the cold gas is evenly distributed in 
a disc, the density threshold can be translated into a critical mass given by

\begin{equation}
m_{\rm crit} = 3.8 \times 10^9
\,\left(\frac{V_{200}}{200 \,\rm km\,s^{-1}}\right)\left( \frac{r_{\rm disc}}{10 \,\rm kpc} \right), 
\end{equation}

\noindent 
where 
$V_{200}$ is the 
circular velocity at the virial radius of the halo
and is used as a proxy for the galactic disc velocity, and $r_{\rm disc}$ is the radius inside which SF occurs. In 
\sag,
we define $r_{\rm disc}={\rm min}(3\,R_{\rm scale}, R_{\rm strip,cold})$, where $R_{\rm scale}$ is the scale radius of the disc (for an exponential density profile, $80$ per cent of the cold gas is contained within
$3\,R_{\rm scale}$), and $R_{\rm strip,cold}$ is the 
galactocentric
distance where the RP exerted over the gas of the disc balances the restoring force of the disc itself (cold gas located beyond this radius is removed by RPS when the condition given by eq.~(\ref{eq:rps}) is fulfilled, as described in the next subsection.)

For a mass of cold gas, $M_{\rm cold}$, the amount exceeding $m_{\rm crit}$ is transformed into stars
at at rate $\dot{M}_{\rm stars}$ according to

\begin{equation}
    \dot{M}_{\rm stars} = \alpha \frac{(M_{\rm cold}-m_{\rm crit})}{\tau_{\rm dyn}},
\end{equation}

\noindent where $\alpha$ is the efficiency of SF, a free parameter of the model ($\alpha \sim 0.04$, see table 1 of \citealt{Cora_2018}), and $\tau_{\rm dyn} = r_{\rm disc}/V_{200}$ is the dynamical timescale of the disc. 

On the other hand, disc instabilities or galaxy mergers trigger starbursts that contribute to the formation of the bulge component and to the growth of central black holes, whose mass can also increase during gas cooling processes taking place once a static hot gas halo has formed around the galaxy \citep{lcp08}. These starbursts are a result of 
the cold gas transfer from the disc to the bulge where it is gradually consumed \citep{Gargiulo_2015}.

\subsection*{Environmental effects}
\label{subsec:rps}

As satellite galaxies orbit within their main host haloes, DM and the baryonic components in their own subhaloes are affected by numerous environmental processes. In particular, we model mass stripping produced by 
RP exerted by 
the ICM\footnote{Although \sag~includes the action of TS, which acts on every component of the galaxy (DM, stars and gas), its implementation produces minor effects on galaxy properties, therefore, we do not consider TS in the current analysis.}, which
is determined by the density of the medium at the satellite position, $\rho_{\rm ICM}$, and the square of the relative velocity between the satellite and the medium, $v^2$.
Specifically, it is defined by
\begin{equation}
    P_{\rm ram} = \rho_{\rm ICM} \, v^2.
    \label{eq:rp}
\end{equation}

Our prescription for RPS of the hot gas halo 
uses the criterion 
applied to a spherical distribution of gas determined by \citet{mccarthy2008} from the results of hydrodynamic simulations,
as implemented by \citet{font2008} in their semi-analytic model.
The gas beyond a satellite-centric radius~$r_\text{sat}$ is removed when the value of RP satisfied the condition
\begin{equation}\label{eq:rpshot}
P_{\rm ram} > \alpha_\text{RP} \frac{G
    M_\text{sat}(r_\text{sat}) \rho_\text{hot}
    (r_\text{sat})}{r_\text{sat}},
\end{equation}
where $\rho_\text{hot}$ is the density of the satellite's hot gas halo, 
for which we adopt an isothermal
density profile, $\rho_\text{hot} \propto r^{-2}$, and the satellite mass $M_{\rm sat}(r_{\rm sat})$ is estimated by integrating this density profile until $r_{\rm sat}$. The geometrical 
constant $\alpha_\text{RP}$ is
chosen to match the results of hydrodynamic simulations; 
we adopt $\alpha_\text{RP}=5$, an intermediate value between those tested by \citet{mccarthy2008}. 

It is worth noting that the cosmic accretion of hot gas is halted in satellite galaxies, but their hot gas halo can be replenished internally by 
the action of SN feedback. On the one hand, this process heats the interstellar medium transporting gas and metals from the cold to the hot gas phase of the same galaxy; in the case of satellites, the reheated mass and associated metals are transferred proportionally to their content of hot gas, while the remains of them are transferred to the corresponding central
galaxy. This scheme of gas replenishment of the hot halo of satellite galaxies applies whenever the ratio
between the hot gas halo and the baryonic mass of the satellite is
larger than a fraction $f_{\rm hot,sat}=0.277$ \citep[see table 1 of][]{Cora_2018}, otherwise, all the reheated mass is transferred to the corresponding central galaxy; note that
the effects of gas cooling 
RPS on satellite galaxies eventually reduce their hot gas halo which can reach a ratio with respect  to the baryonic mass lower than $f_{\rm hot,sat}$.
On the other hand, SN feedback ejects part of the hot gas out of the (sub)halo which is reincorporated latter; this process in implemented to avoid an
excess of stellar mass at high redshifts (for more details, see sections 3.4 and 4 of \citealt{Cora_2018}).

We assume that the hot gas halo of a satellite acts as a shield against the action of RP on the cold gas present in the disc as long as the ratio
between the hot gas halo and the baryonic mass of the galaxy is larger than $0.1$.
When such a ratio becomes lower than this value, the hot gas halo is considered depleted and RP can strip the cold gas disc according to the condition given by \citet{gg72}, as implemented by \citet{tecce10}.
In this case, RP must exceed the restoring force per unit area due to the gravity of the disc. i.e.
\begin{equation}\label{eq:rps}
  P_{\rm ram} > 2\pi G \Sigma_\text{disc}(R)
  \Sigma_\text{cold}(R),
\end{equation}
\noindent to remove cold gas from the disc; here $\Sigma_\text{disc}$ and $\Sigma_\text{cold}$ are the surface densities of the
galactic disc (stars plus cold gas) and of the cold gas disc, respectively.
They are modelled with
exponential profiles with the same scale-length.

The values of RP, $P_{\rm ram}$, defined in eq.~(\ref{eq:rp}) and involved in eqs.~(\ref{eq:rpshot}) and (\ref{eq:rps}), are obtained from a new analytic fitting profile that model the RP exerted over satellite galaxies in different environments and epochs \citep{vega2022}.
The gas removed by RPS of both the hot and cold gas components are incorporated to the hot has halo of the central galaxy, i.e. the stripped mass contributes to the increment in mass and metals of the ICM.

\subsection*{Orbital evolution of orphan galaxies}

The information about the orbital evolution of satellites is crucial for the adequate estimation of RPS,
since the value of RP defined in eq.~(\ref{eq:rp}) 
depends on both the cluster-centric distance and velocity of the satellite galaxy
through 
the quantities involved, 
$\rho_{\rm ICM}$ and $v$, respectively.
Our galaxy catalogue considers orphan satellites whose positions and velocities are consistently
calculated by tracking the orbital evolution of unresolved subhaloes in a pre-processing step, before applying \sag~to the DM-only simulations.  
This is an important advantage with respect to the treatments generally applied in  semi-analytic models, which follow the orbital evolution of orphan galaxies in a simplified way, like estimating the moment in which they merge with the central galaxy of their host structure by considering the orbital decay time-scale due to dynamical friction 
\citep{Chandrasekhar_DynamicalFriction_1943A, Boylan-Kolchin_MergingTimescale_2008}.

In our implementation,
when a subhalo is not longer detected, its last known position, velocity, and virial mass are taken as initial conditions to integrate its  orbit numerically in the potential well of the host halo. We use an orbital evolution model that considers dynamical friction and mass-loss by TS. The general description of this model is given by \citet{Delfino_2022}. It is worth noting that, in their version of the orbital evolution code, the mass profile of both the host halo and unresolved subhaloes are modelled by a NFW profile, while in the previous version used in this work, we adopt an isothermal sphere.

The integration of the orbit of an orphan galaxy is carried out until it merges with its central galaxy. We consider that the merger event occurs when the halo-centric distance of the orphan satellite becomes smaller than 10 per cent of the virial radius of the host halo, taken as an estimation of the radius of a central galaxy\footnote{At this stage, we do not have information on the properties of galaxies that will populate these haloes because the orbital integration is done before applying \sag~to the DM-only simulation.}.

The DM simulation has a temporal resolution of $\sim 430\,{\rm Myrs}$ at ${z\sim 0}$ and of $\sim 140 \, {\rm Myrs}$ at ${z\sim 2}$, which is a reasonable time resolution to properly track the orbital evolution of galaxies within the scope of this work 
(see Fig.~\ref{fig:distance-evolution} and the corresponding discussion in Sec.~\ref{sec:subpopulations}).
We highlight that, in this model, some orphan galaxies bound
to the cluster have apocentres that lie in the outskirts of the cluster, i.e. the region that extends beyond \rdos. This is further detailed in Sec.~\ref{subsec:distribution}.

\begin{figure}
    \centering
    \includegraphics[width=\columnwidth]{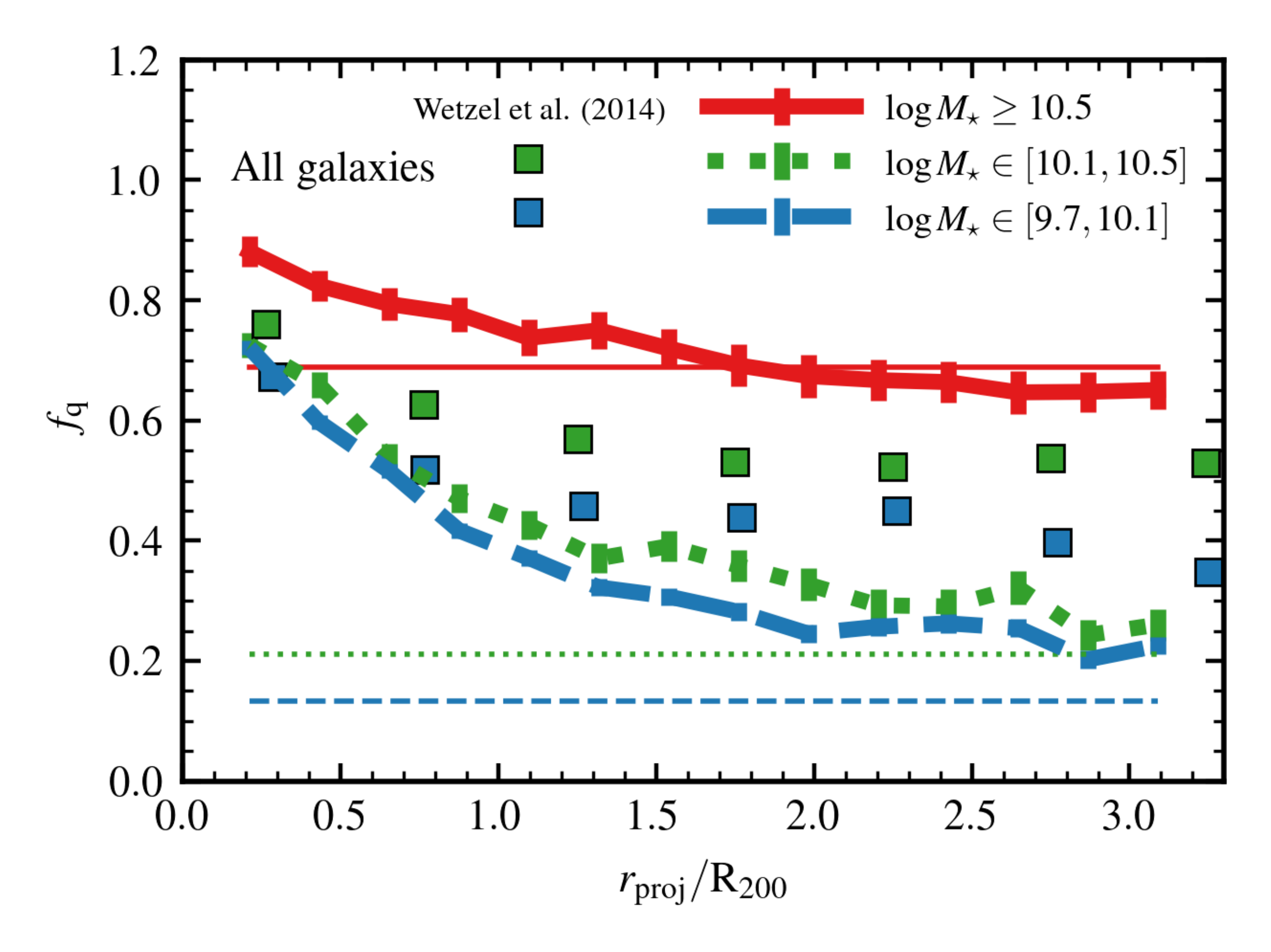}
\caption{Radial dependence of the fraction of quenched galaxies ($f_{\rm q}$) 
for different stellar mass bins at $z=0$. We build a unique sample by stacking the galaxy population in and around the 102 relaxed galaxy clusters, and separate them by stellar mass, as indicated in the legend. Coloured thick lines represent the dependence of $f_{\rm q}$ 
on normalized cluster-centric distance in projected space, while coloured thin horizontal lines represent the average throughout the complete 
\mdpl~simulation. Error bars show $68$ per cent confidence interval of the average for a beta distribution \citep{Cameron11}. 
The fraction $f_{\rm q}$ for the two lower stellar mass bins are compared with data from the SDSS \citep[][filled squares]{Wetzel_2014}.
} 
    \label{fig:fq-mstar-r-All-Wetzel14}
\end{figure}

\subsection{Fraction of quenched galaxies as a function of clustercentric distance}

The study of the gradient of the radial distributions
of the fraction of quenched galaxies broadly relates the evolutionary state of a galaxy population, the environment in which galaxies reside and their SF activity. While there is a correlation between satellite cluster-centric distance and time since infall onto the cluster \citep{delucia2012,oman13,smith2019,mostoghiu21},
galaxies located outside \rdos~are part of a more heterogeneous sample (see Sec.~\ref{sec:subpopulations}). However, radial gradients have been effective in showing that the effects of environment persist beyond the virial radius of clusters \citep{Wetzel12, cohen2014,haines15,adhikari2021, lacerna2022}.

In \citet{Cora_2018}, we show that the \sag~model successfully
reproduces the dependence of the fraction of quenched galaxies ($f_{\rm q}$) on cluster-centric distance for different stellar and halo mass ranges in the complete \mdpl~simulation (see their fig. 12), although $f_{\rm q}$ results slighlty underestimated for galaxies with $M_\star = 10^{9.7}-10^{10.5}\, {\rm M_\odot}$ in haloes of $M_{\rm halo} = 10^{14.1}-10^{15}\, {\rm M_\odot}$.
Throughout this work, we consider a galaxy as quenched if its 
sSFR satisfies that
${\rm log(sSFR [yr^{-1}])} < -10.7$, given that this threshold separates the bimodal distribution of the $\rm sSFR$ in the current version of \sag, \citep[see][for further details]{Cora_2018}. 
In Fig.~\ref{fig:fq-mstar-r-All-Wetzel14}, we show 
$f_{\rm q}$ 
as a function of projected distance to the cluster centre (defined by the location of the central galaxy of the cluster) normalized with the radius \rdos~of each cluster. Model galaxies (different lines with error bars) are binned in stellar mass according to the selection of observed galaxies analysed by \cite{Wetzel_2014}. Thin horizontal lines represent the average of $f_{\rm q}$ throughout the complete \mdpl~simulation. 
Observations by \citet{Wetzel_2014} (depicted by filled squares) show a sharp increase of $f_{\rm q}$ at 
a projected distance $ r\lesssim 1.5-2\, R_{200}$, 
although they report an enhancement up to $6\,R_{200}$ with respect to the cosmic average at each stellar mass considered (see their fig. 1). 
Our model reproduces the enhancement of $f_{\rm q}$ at $r\lesssim 2\, R_{200}$ for the lower stellar mass bins, in consistency with the analysis done in \citet{Cora_2018}, indicating that, as galaxies approach regions with progressively higher densities of galaxies,
its SF
activity is largely affected.
Although $f_{\rm q}$ results underestimated with respect to observations, specially at $\rm r > 1 \, R_{200}$, the relative impact of the environment manifested on the trend is very similar to what observations show. The discrepancy is likely to arise from the fact that, in the current version of \sag, galaxies with ${\rm log(}M_\star [ {\rm M_\odot} ]) \in [10.1,10.5]$ tend to be more star-forming than observed, independently of the environment 
where they reside \citep[see fig. 11 of ][]{Cora_2018}.
Regarding the lowest mass bin, the underestimation might indicate that pre-processing effects are not being strong enough, or that environmental effects inside the cluster are insufficient to affect the SF of the current backsplash population. In fact, \citet{sarron2019} shows that the passive fraction result higher in galaxies located in filaments than in istropically selected regions around clusters, which highlights the role of filaments in quenching galaxies before they reach the cluster virial radius. \sag~do not include a specific treatment of environmental effects associated to the gas present in filaments, which could also be significant in the discrepancy found between the model and the observations (see Sec.~\ref{sec:transformation} for further discussion).

The RPS model implemented in \sag~can affect 
cluster satellites
located outside \rdos, although the values of the RP are several orders of magnitude lower than inside \rdos; however, only $\sim 10$ per cent of the galaxies located in the outskirts are classified as cluster satellites (most of them are orphans). 
For 
the two lowest stellar
mass bins, $f_{\rm q}$ in real space results slightly higher than in projected space (not shown here) because, in the latter, 
some galaxies that are located further from the cluster (and hence tends to be more star-forming) appears artificially closer to the cluster centre. 

For completeness, we decide to include in Fig.~\ref{fig:fq-mstar-r-All-Wetzel14} our simulated high-mass galaxies, which are mainly sensitive to mass quenching and, as we can see, the environment generates a moderate increase in the 
fraction of quenched galaxies
from $f_{\rm q}\sim 0.7$ at $1.5-3$\rdos~to $f_{\rm q}\sim 0.95$ at $0.2$\rdos. 
Galaxies with ${\rm log(}M_\star [{\rm M}_\odot]) \in [9,9.7]$, which are characterized by a high proportion of orphans, are not included in the plot, but we mention that both their trend and normalization resemble those of ${\rm log(M}_\star [{\rm M}_\odot]) \in [9.7,10.1]$, 
indicating
that the evolution of their SF and the impact of environmental effects over this population are well traced.

\section{Galaxy population in and around clusters}
\label{sec:populations}

The aim of this work is to understand the physical reasons that regulate the SF 
of galaxies in and around galaxy clusters.
We consider \rdos~as the radius of the spherical limiting boundary between these two regions; the outskirts is considered to be comprised between \rdos~and $3\,R_{200}$.

In this section, we first analyze the radial distribution of all galaxies within $3\,R_{200}$ 
of the cluster centre,
according to their classification in central and satellite galaxies,
and then we define galaxy samples according to their current locations and orbital history.

\begin{figure}
    \centering
    \includegraphics[width=\columnwidth]{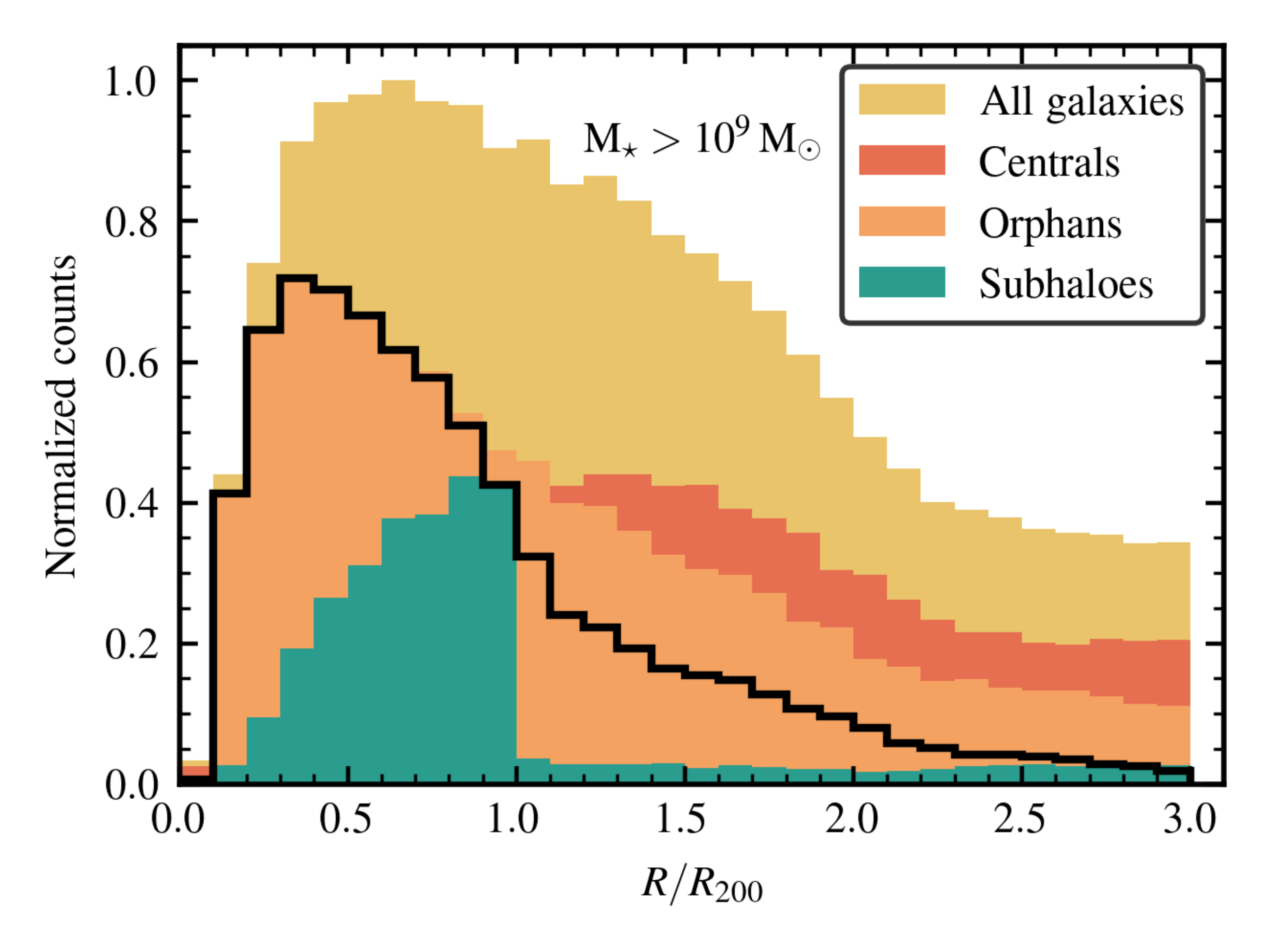}
    \caption{Radial distribution of galaxies in and around clusters according to the distance to the central galaxy of the corresponding galaxy cluster, separated by type: centrals, satellites that keep their DM subhalo, and orphan satellites. 
    The latter are discriminated between orphan satellites
    gravitationally bound
    to the most massive DM halo (orphans of the cluster halo, solid black line) and orphan satellites of the neighbouring centrals. Subhaloes beyond \rdos~are too few to be discriminated in this way and, basically, there are no subhaloes that belong to the cluster outside \rdos~(see Table~\ref{Table:Percentage2}). The sharp drop of subhalo counts at \rdos~reflects the fact that the vast majority of the galaxies with DM-haloes outside \rdos~are centrals.
    The galaxy counts are normalized 
    with the maximum number of galaxies from all bins (where the distribution reaches $1$, given by $N_{\rm max}$; here $N_{\rm max}=4015$).
    } 
    \label{fig:histo-r-Type1Type2}
\end{figure}

\begin{table}
    \centering
    \begin{tabular}{l | c |  c} 
    
        Population & $\rm r\,\leq \,R_{200}$ & $\rm R_{200}\,<\,r\,\leq\,3\, R_{200}$ \\
        \hline
        Subhaloes & 0.24 & 0.04 ($\sim$ 0) \\
        Orphans & 0.76 & 0.41 (0.19) \\
        Centrals & - & 0.55 \\
        \hline
        Recent infallers & 0.27 & - \\
        Ancient infallers & 0.73 & - \\
        Backsplash & - & 0.48 \\
        Neighbouring & - & 0.52 \\
         
         \bottomrule
    \end{tabular}

    \caption{Fraction of different sets of galaxies populating the inner region (${\rm r}\leq R_{200}$) and outskirts ($R_{200} <r\leq 3\, R_{200}$) of the sample of the 102 relaxed galaxy clusters. 
    Galaxies are selected according to their stellar mass content at $z=0$; they have $ M_\star \geq 10^{9}\, {\rm M}_\odot$. 
    In the upper part of the table, the whole galaxy population is discriminated in central galaxies, 
    satellites that keep their DM
    haloes and orphans.
    In the case of 
    subhaloes and orphans
    located in the outskirts, the first number 
    involves
    satellites of the clusters and of neighbouring centrals, and the second (indicated between brackets) only considers satellites of the cluster identified in the region. In the lower part of the table, the whole galaxy population is  discriminated according to galaxy types defined from their orbital evolution and current cluster-centric distance: recent infallers (RIN), ancient infallers (AIN), backsplash galaxies (BS) and neighbouring galaxies.
    }
    \label{Table_Percentage}
\end{table}

\begin{table}
    \centering
    \begin{tabular}{l | c |  c} 
    
        Type & $\rm N_{type }/N_{backsplash}$ & $\rm N_{type}/N_{neighbouring}$ \\
        \hline
        Centrals & 0.47 & 0.62 \\
        Subhaloes & 0.02 & 0.07 \\
        Orphans & 0.51 & 0.31 \\
       
         \bottomrule
    \end{tabular}

    \caption{
    Composition in terms of central galaxies, satellite galaxies with DM haloes and orphans of the populations of backsplash and neighbouring galaxies
    located in the outskirts ($R_{200}<r\leq 3\, R_{200}$) of the $102$ relaxed galaxy clusters. Galaxies are selected according to their stellar mass content at $z=0$, and have $M_\star \geq 10^{9}\, M_\odot$.
    }
    \label{Table:Percentage2}
\end{table}

\subsection{Radial distribution of central galaxies, satellites with DM substructures and orphan satellites}
\label{subsec:distribution}

We evaluate the radial distribution
of galaxies in and around clusters.
The whole galaxy population within $3\,R_{200}$ is separated in central galaxies of surrounding DM haloes and satellite galaxies of either the main clusters or the haloes located in their outskirts. In turn, satellites are classified as galaxies with DM
haloes or orphans depending on whether they keep their DM halo or not. From now on, we will refer to satellites that keep their DM halo as `subhaloes', for simplicity.
The relative contribution of 
each type of galaxy
to the whole galaxy population as a function of their cluster-centric distance, i.e. the distance to the centre of the most massive
halo of each selected region, can be appreciated in Fig.~\ref{fig:histo-r-Type1Type2}.

We first focus on satellite galaxies orbiting within \rdos.
On average, $24$ per cent of them are 
subhaloes while the remaining $76$ per cent are orphans. 
As can be seen, the population of orphan satellites gains relevance for lower cluster-centric distances. This is consistent with the fact that these galaxies have been accreted much earlier than  subhaloes ($\sim 70$ per cent of current orphans have infall times\footnote{See Sec.~\ref{sec:subpopulations} for details on the definition of infall and first infall.}
greater than $5 \, {\rm Gyr}$, while $\sim 60$ per cent of subhaloes have infall times lower than $3\, {\rm Gyr}$)
and have had enough time to be affected by dynamical friction effects bringing them closer to the cluster centre.

Fig.~\ref{fig:histo-r-Type1Type2} also shows that the population of galaxies  contained in the volume comprised between the spheres of radii \rdos~and $3\,R_{200}$ is dominated by central galaxies, referred to as neighbouring centrals, followed by orphan satellites. Considering the galaxy population outside \rdos, we find that
neighboring centrals represent $\sim 55$ per cent of the whole galaxy population in this region, while orphan satellites contribute with $\sim 41$ per cent. The remaining small percentage of $\sim 4$ per cent corresponds to subhaloes. In turn, both types of satellites can be either satellites of the clusters or of their neighbouring centrals. In the case of orphan satellites located in the outer region, around half of them is gravitationally bound to the clusters.
There are almost no galaxies with subhaloes gravitationally bound to the clusters further than \rdos.
All these percentages are presented in
the upper part of Table~\ref{Table_Percentage} in order to summarize the general situation.

\subsection{Classification of galaxies according to their orbits and current locations}
\label{sec:subpopulations}

The need for an interpretation of the SF quenching of galaxies in and around clusters
in terms of physical processes that affect 
them
calls for a classification according to their current cluster-centric distance and their orbital evolution.
There are many evolutionary pathways that a galaxy can follow to constitute the satellite population, or the outskirt population of a cluster.
We define a cluster galaxy as a galaxy located within the radius \rdos~of the cluster. Note that a galaxy that is gravitationally bound to the main host halo of a galaxy cluster 
can be found outside \rdos.
However, such a galaxy is considered as being part of the outskirts of the clusters with the aim of constructing subsamples of galaxies that are mutually exclusive.

Although the incorporation of galaxies into the cluster is a continuous process, we identify two stages in the build up of the current satellite population. 
On the one hand, many satellites are accreted during the early stage of cluster evolution (between $6$ and $10$ Gyr ago) and still remain inside the cluster by the present time. On the other hand, current satellites can be recently accreted (less than $2$ Gyr ago).
Galaxies around clusters can be either galaxies that have been inside \rdos~and now are outside (known as backsplash galaxies), or can be neighbouring centrals with their own satellites, which have never been inside the cluster.

In this work, we define the time of infall as the moment when a galaxy crosses \rdos~of the main progenitor of the host halo inward for the first time. It is worth noting that this definition differs from the one adopted in \citet{Cora_2018}, where they have used the moment when a galaxy becomes a satellite of any halo (which may or may not coincide with
the main progenitor of the cluster) for the first time. The latter definition includes all physical processes associated to being a satellite, while the former definition (considered in the current work) allows us to discriminate the impact of the environment of the cluster from the secular evolution or pre-processing effects that take place before infall.

\begin{figure}
    \centering
    \includegraphics[width=\columnwidth]{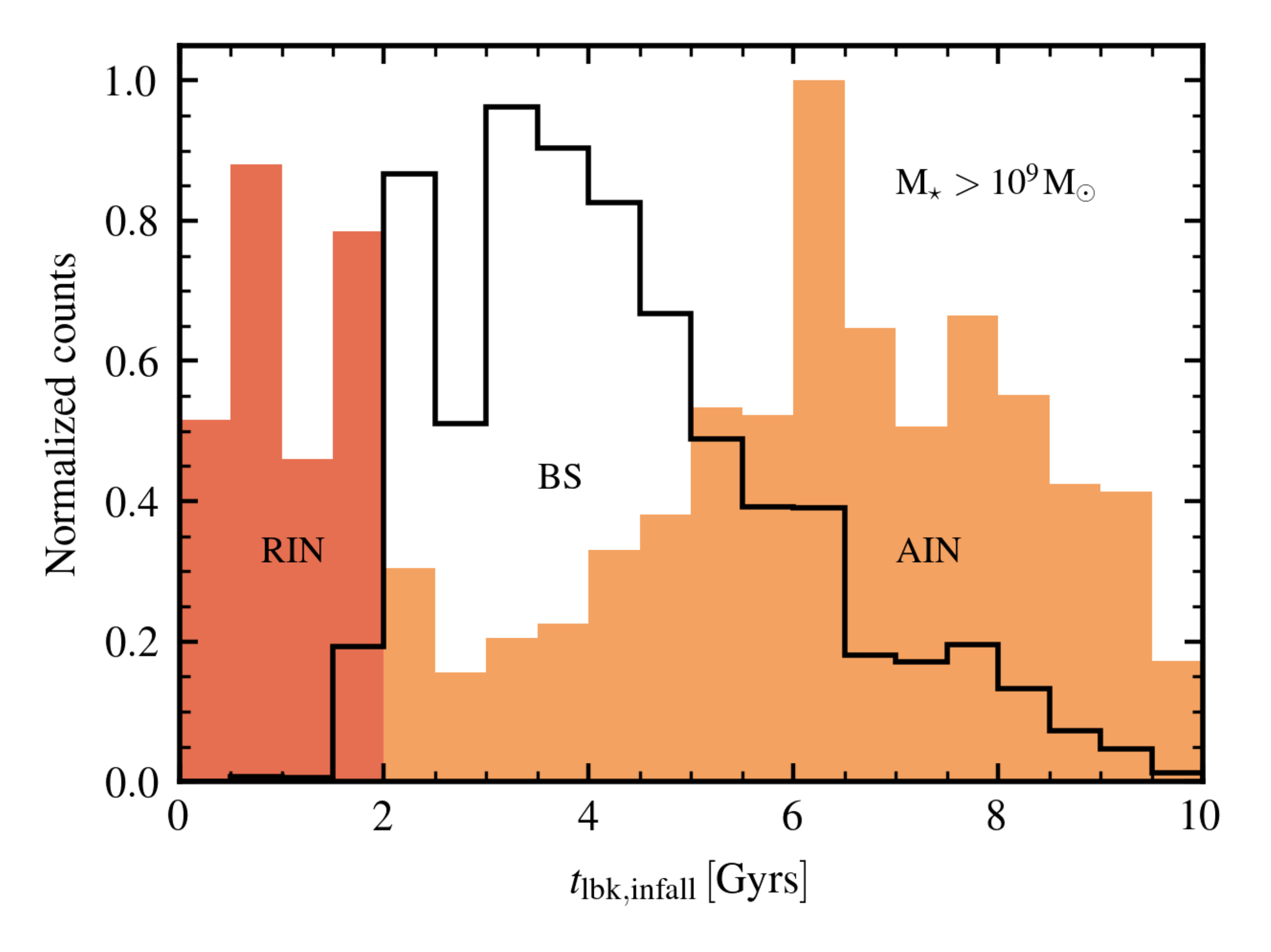}
    \caption{Histograms 
    of lookback infall time, $t_{\rm lbk,infall}$, for ancient infallers (AIN, orange histogram), recent infallers (RIN, red histogram) and backsplash galaxies (BS, histogram depicted by a black solid line). This time is defined as the time since a galaxy have crossed \rdos~of the main progenitor of the cluster halo.
    The galaxy counts are normalized with the maximum number of galaxies from all bins achieved by any of the galaxy populations (in this case, AIN, where the distribution reaches $1$), given by $N_{\rm max}$; here $\rm N_{max} = 4587$. 
    }
    \label{fig:histo-time-inside}
\end{figure}

Fig.~\ref{fig:fq-mstar-r-All-Wetzel14} shows broadly the impact of environmental processes over the $\rm SFR$ of galaxies within $3$\rdos.
In order to link 
the evolution of their $\rm SFR$ and other properties
with their orbital history,
we classify galaxies within the clusters and in their surroundings in four subpopulations:

\begin{itemize}
    \item Ancient infallers (AIN): satellite galaxies located within $R_{200}$ of the cluster at $z=0$ (cluster galaxies) that have crossed this radius for the first time more than $2$ Gyr ago, i.e. in a lookback time $t_{\rm lbk,infall}> 2\,{\rm Gyr}$.
    \item Recent infallers (RIN): cluster galaxies with $t_{\rm lbk,infall} \leq 2\,{\rm Gyr}$.  
    \item Backsplash galaxies (BS): galaxies with a $z=0$ cluster-centric distance larger than \rdos, but that have been inside the cluster in the past at some point of their orbital evolution. 
    \item Neighboring galaxies (NB): Galaxies located further than \rdos~from the cluster centre and that have never been within the cluster.
\end{itemize}

\citet{haggar2020} find that, from the galaxies located between $R_{200}$ and $2\,R_{200}$, the fraction of galaxies that are backsplash is higher for the sample of relaxed clusters ($\sim 70$ per cent) than for the non-relaxed one ($\sim 40$ per cent) at ${z=0}$; the evolution of the backsplash fraction for these two samples somewhat diverges after $z\sim 0.4$, around $4$ Gyr ago \citep[see fig. 5 of][]{haggar2020}.
Note that NB constitute a sample formed by neighbouring central galaxies and their corresponding satellites, and is considered only for completeness in order to have a clearer global view of the radial distribution of the passive galaxies (see Fig.~\ref{fig:passive-radial} in Section~\ref{sec:quenched_fractions}); the rest of the analysis carried out in this work will be focused on AIN, RIN and BS.

In Fig.~\ref{fig:histo-time-inside}, we present the normalized distribution of the lookback
time since crossing the radius \rdos~of the 
main progenitor of the cluster halo
for all 
galaxies (black solid line), RIN (red histogram) and AIN (orange histogram). The majority of AIN have crossed \rdos~more than $\sim 5 \, {\rm Gyr}$ ago, and the current BS population has infall times between $\sim 2$ and $\sim 6\,{\rm Gyr}$ ago. The typical crossing timescale of our cluster sample is of $\sim 2-4\,{\rm Gyr}$, which is consistent with the fact that the majority of the galaxies that crossed \rdos~less than $2\,{\rm Gyr}$ ago are classified as RIN. This supports our choice of 
an infall lookback time of $2\,{\rm Gyr}$
to separate RIN from AIN, also considered in \citet{smith2019} and \citet{delosrios21}.
This classification is robust, as considering a larger lookback time of $t_{\rm lbt,infall} \sim 3-4\,{\rm Gyr}$ does not affect significantly our results on the fraction of quenched galaxies or the gas fraction presented in Figs.~\ref{fig:fq-mstar_AIN-RIN-BS} and \ref{fig:fhot-mstar_AIN-RIN-BS}, respectively.
When we analyze the moment of infall and exit of BS, we find that, in general, BS spend less than $2\,{\rm Gyr}$ inside the cluster (with a peak on the distribution of $\sim 1.5 \,{\rm Gyr}$, not shown here), which restricts the impact of environmental effects on this population. 

AIN represent $\sim 73$ per cent of the current satellite population located inside \rdos, and are mainly orphan galaxies, while the remaining $\sim 27$ per cent is constituted by RIN (see lower part of Table~\ref{Table_Percentage}). By definition, AIN can cross \rdos~several times along their orbital evolution and currently lie inside \rdos. 
BS constitute $\sim 48$ per cent of the galaxies located between $1-3\, R_{200}$, in agreement with \citet{gill05} and \citet{Wetzel_2014}. \citet{haggar2020} show that, at $z=0$, ${\sim 70}$ per cent of the galaxies located between $R_{200}<r<2\, R_{200}$ are 
BS in the galaxy cluster sample of
\textsc{The Three Hundred} project generated with hydrodynamical simulations, although the stellar and halo mass limit of their sample is slightly different with respect to
ours, and they do not include orphan galaxies.
BS include galaxies that have crossed \rdos~several times in their way in and out the cluster, although they represent a minor population; BS sample is dominated by galaxies that have crossed \rdos~only twice ($83$ per cent), in consistency with \citet{Wetzel_2014} and \citet[][$90$ per cent between \rdos~and $2$\rdos]{haggar2020}. 

Our BS sample partially match the `ejected' galaxies analysed by \cite{Wetzel_2014}, 
as in their sample they also include galaxies that were ejected from neighbouring host haloes. 
In fact, from their fig. 2, we infer that our BS
sample (although we include orphans) resembles the distribution of their satellites that were ejected only from the host haloes on which the profiles were centred, which they rapidly decay at $\sim 3 \, R_{\rm 200c}$.

\begin{figure}
    \centering
    \includegraphics[width=\columnwidth]{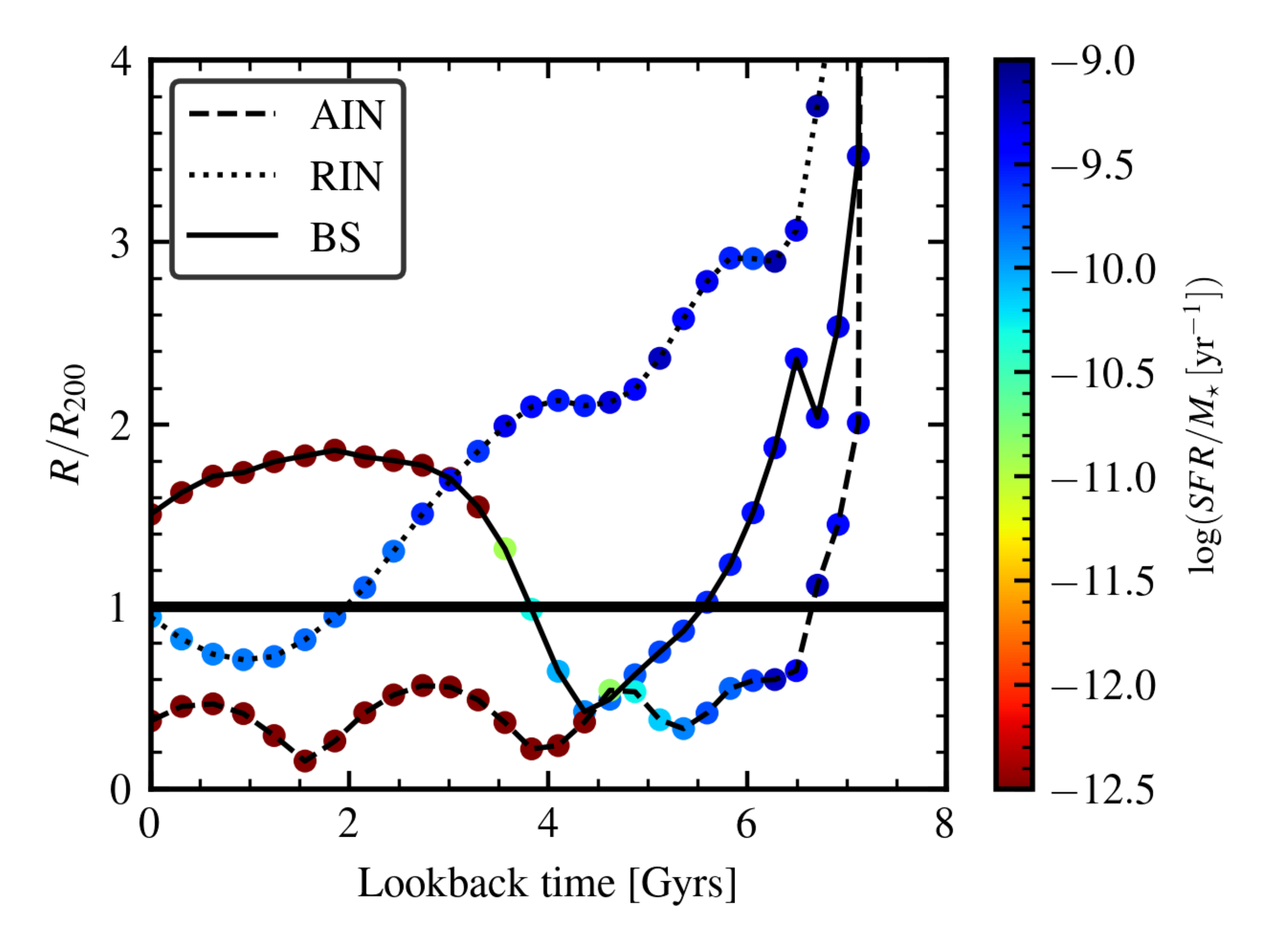}
    \caption{Normalized distance to the cluster centre as a function of lookback time for three randomly selected galaxies for our sample. We select an 
    AIN dashed), a RIN (dotted) and a BS
    (solid line). Each point is colour-coded according to the sSFR of the galaxy. }
    \label{fig:distance-evolution}
\end{figure}

As an example,  Fig.~\ref{fig:distance-evolution} shows the time evolution of the distance to the cluster centre of a randomly selected galaxy of each of the first three sub-samples 
(AIN, RIN, BS)
characterized by the corresponding values of the sSFR throughout its orbit. 
The AIN crosses \rdos~$\sim 7 \,{\rm Gyr}$ ago, and it becomes
passive $\sim 2.5\,{\rm Gyr}$ after infall, after its first pericentric passage and before experiencing the second one.
BS only spend $\sim 1\,{\rm Gyr}$ inside the cluster, and remain star-forming while being located within the cluster; their SF is quenched outside $R_{200}$. The closest distance reached by RIN is $\sim 0.8 \, R_{200}$, and they remain forming stars till the present.

\section{Ancient and recent infallers, backsplash galaxies}
\label{sec:AINRINBS}

With the aim of understanding the physical reasons involved in the quenching of SF of galaxies within clusters and in their outskirts, we analyse
the evolution of the fraction of quenched galaxies, 
the correlation between the gas content and phase-space position, 
the evolution of the 
hot and cold gas content of these galaxies, and the connection of their SF history with pericentric passages,
focusing on the samples of AIN, RIN and BS.

\subsection{Fractions of quenched galaxies}
\label{sec:quenched_fractions}

\begin{figure}
    \centering
    \includegraphics[width=\columnwidth]{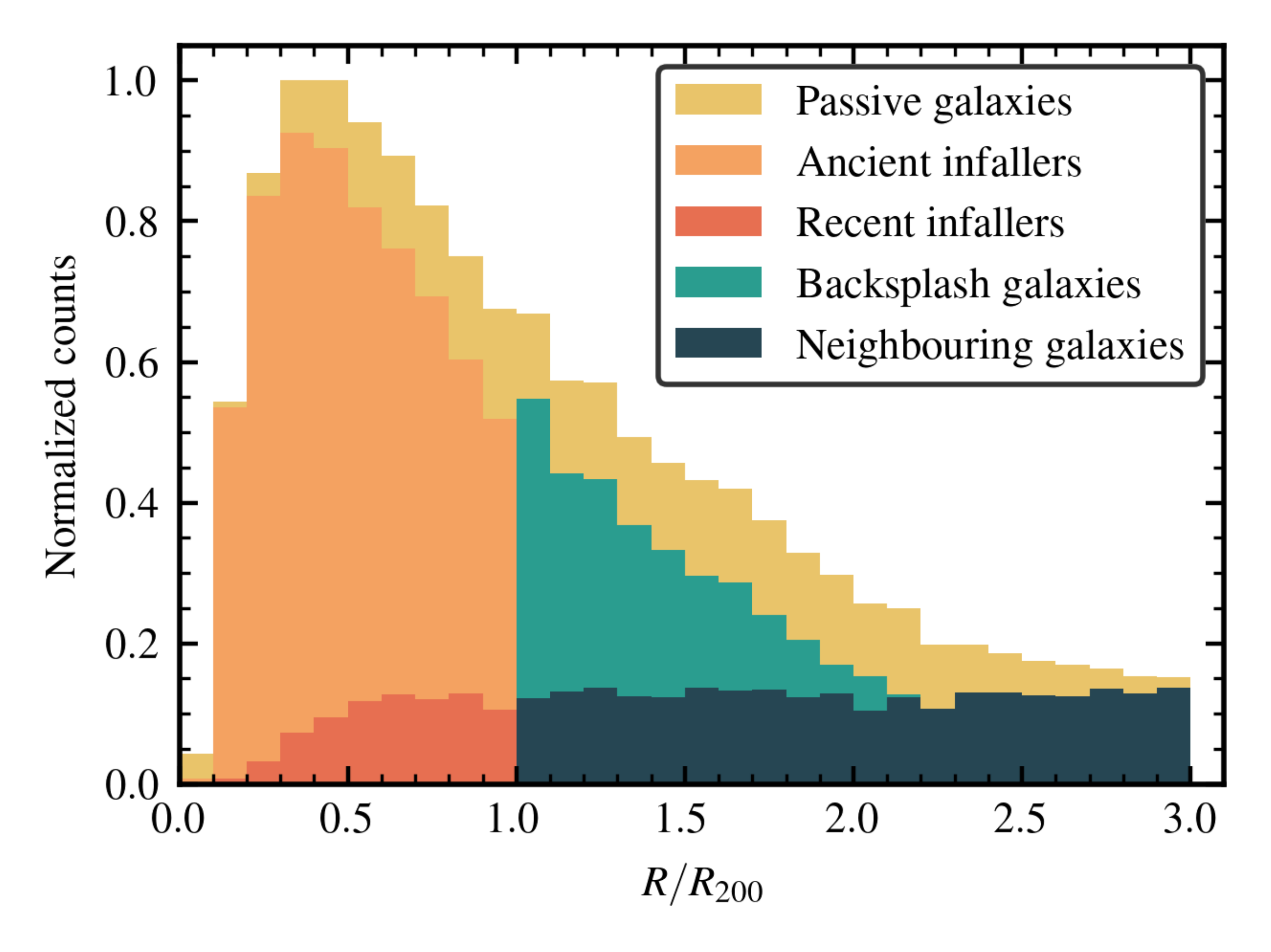}
    \caption{Radial distribution of all the passive galaxies (${\rm sSFR} < 10^{-10.7} {\rm yr^{-1}}$)
    in and around the clusters in our sample, separated in  
    different populations according to their orbital history and current cluster-centric distance.
    The galaxy counts are normalized 
    with the maximum number of galaxies from all bins (where the distribution reaches $1$, given by $N_{\rm max}$; here $ N_{\rm max}= 2927$).
    }
    \label{fig:passive-radial}
\end{figure}

\begin{figure*}
    \centering
    \includegraphics[width=\textwidth]{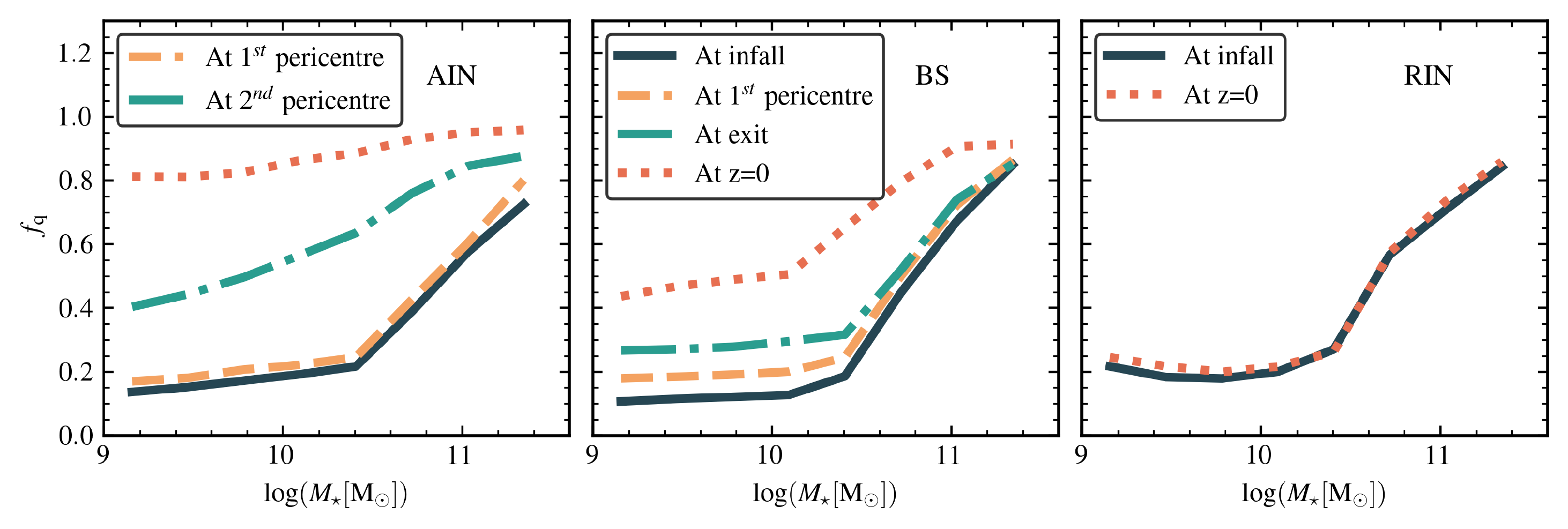}
    \caption{Fraction of quenched galaxies, $f_{\rm q}$,  as a function of stellar mass for ancient infallers (left panel), backsplash galaxies (middle panel) and recent infallers (right panel) at different moments of their life time: infall time (solid dark line), at the moments of first and second pericentric passage (orange dashed and green dashed-dotted lines, respectively) and at $z=0$ (red dotted line). In the case of BS, we show the moment they cross $\rm R_{200}$ outwards (green dash-dotted line), as the vast majority of them only have one pericentre passage.
    }
    \label{fig:fq-mstar_AIN-RIN-BS}
\end{figure*}

Since we are interested in the origin of the population of passive galaxies, we first analyse how the different galaxy samples defined in Sec.~\ref{sec:subpopulations} constitute the passive population in and around clusters.
Fig.~\ref{fig:passive-radial} shows a radial distribution of passive galaxies consistent with the radial dependence of the fraction of quenched galaxies presented in Fig.~\ref{fig:fq-mstar-r-All-Wetzel14}, i.e. lower numbers of passive galaxies for larger cluster-centric distances.
For $r<R_{200}$, passive AIN clearly outnumber passive RIN, while 
in the outskirts of the clusters
BS dominate the passive population up to $r\sim 2.2  R_{200}$.
Considering the values presented in Table~\ref{Table:Percentage2}, 
central galaxies represent more than half of 
the neighbouring galaxies, 
which indicates that the suppression of SF of such population is more related to mass quenching processes rather than to environmental ones, at least in the context of our model. 
In general, quenched low-mass central galaxies are much closer to massive haloes than star-forming central galaxies of the same mass, by a factor of $\sim 5$, a trend that is found in \sag~and also in the \textsc{Illustris} hydrodynamical simulation \citep{lacerna2022}.

We explore the dependence of the fraction of passive or quenched galaxies, $f_{\rm q}$, of each galaxy population on stellar mass at $z=0$, at different moments of their evolution: at infall, at the moments of first and second pericentric passages, and at $z=0$.
To determine these moments, we follow the evolution of the distance of each galaxy to the cluster centre, and calculate the moment when the galaxy crosses \rdos~(infall time), and when the distance reaches a local minimum inside \rdos~(for first, second or subsequent pericentric passages).
On the left panel of Fig.~\ref{fig:fq-mstar_AIN-RIN-BS}, we can appreciate the strong evolution of  $f_{\rm q}$ with time, specially for low-mass ($M_\star \leq 10^{10.5}\, {\rm M_\odot}$) AIN, from $\sim 0.2$ at infall (dark solid line), reaching $\sim 0.3-0.4$ at the moment of their second pericentric passage (green dashed-dotted line), 
and finally to $\sim 0.8-0.85$ at $z=0$ (red dotted line). 
It is clear that a long time inside the cluster,
under the action of RP,
the occurrence of several pericentric passages (see Sec.~\ref{subsec:pericentre}) where the environmental effects are stronger, and the natural gas consumption into stars have high impact on the SFR of this population. 
In contrast, high-mass ($ M_\star > 10^{10.5} \, {\rm M_\odot}$) AIN show less evolution of their $f_{\rm q}$
as they tend to be 
more quenched than low-mass galaxies at their infall times.
This is because current high-mass galaxies form their stars earlier than low-mass ones (consuming their gas reservoirs) and AGN feedback prevents further gas cooling fueling their cold gas disc
\citep{cora2019}, thus interrupting the baryon cycle.

For BS (middle panel of Fig.~\ref{fig:fq-mstar_AIN-RIN-BS}), we identify three other timestamps during their orbit: the moment of infall, the moment of first pericentric passage and the most recent moment of crossing \rdos~outward (taking into account that a small fraction of BS have crossed \rdos~inward and outward several times). 
For low-mass BS, 
$f_{\rm q}$
increases from $\sim 0.1$ at infall to $\sim 0.3$ at exit 
showing the modest impact of the environmental effects on their SF activity. 
At their first pericentric passage (although, in general, BS have only one), BS
can reach a cluster-centric distance of $\sim 0.2-0.3 \, R_{200}$, where RPS is extreme, but it seems to be 
not enough to foster an efficient quenching of the 
SF some time later; however, taking into account that only around half of the low-mass BS are passive at $z=0$,
a deeper analysis separating passive and star-forming BS shows a more complex scenario (see discussion related to Fig.~\ref{fig:fhot-mstar_AIN-RIN-BS}).
Regarding the high-mass BS, the fraction $f_{\rm q}$
shows
little evolution while they orbit inside the cluster. 
However, at $z=0$, we find that this fraction has increased
up to $f_{\rm q}\sim 0.6$ for $M_\star=10^{10.5}\, {\rm M_\odot}$, and $f_{\rm q}\sim 0.9$ for $M_\star \geq 10^{11}\, {\rm M_\odot}$.

On the other hand, RIN present
a negligible evolution of their $f_{\rm q}$ between infall time and $z=0$, regardless of their $z=0$ stellar mass content (right panel of Fig.~\ref{fig:fq-mstar_AIN-RIN-BS}). 
As only $\sim 50$ per cent of RIN experience a pericentric passage, and the time evolution of $f_{\rm q}$ is negligible,
we decide not to include 
timestamp corresponding to such event
in the figure.
The behaviour of RIN shows that $2 \rm \, Gyr$ is not a long enough period of time for environmental effects (at least those included in our model, as RPS of both hot and cold gas; see Sec.~\ref{sec:model}) to efficiently affect the SF activity of a satellite population, even in high-density environments.

\begin{figure*}
    \centering
    \includegraphics[width=0.8\textwidth]{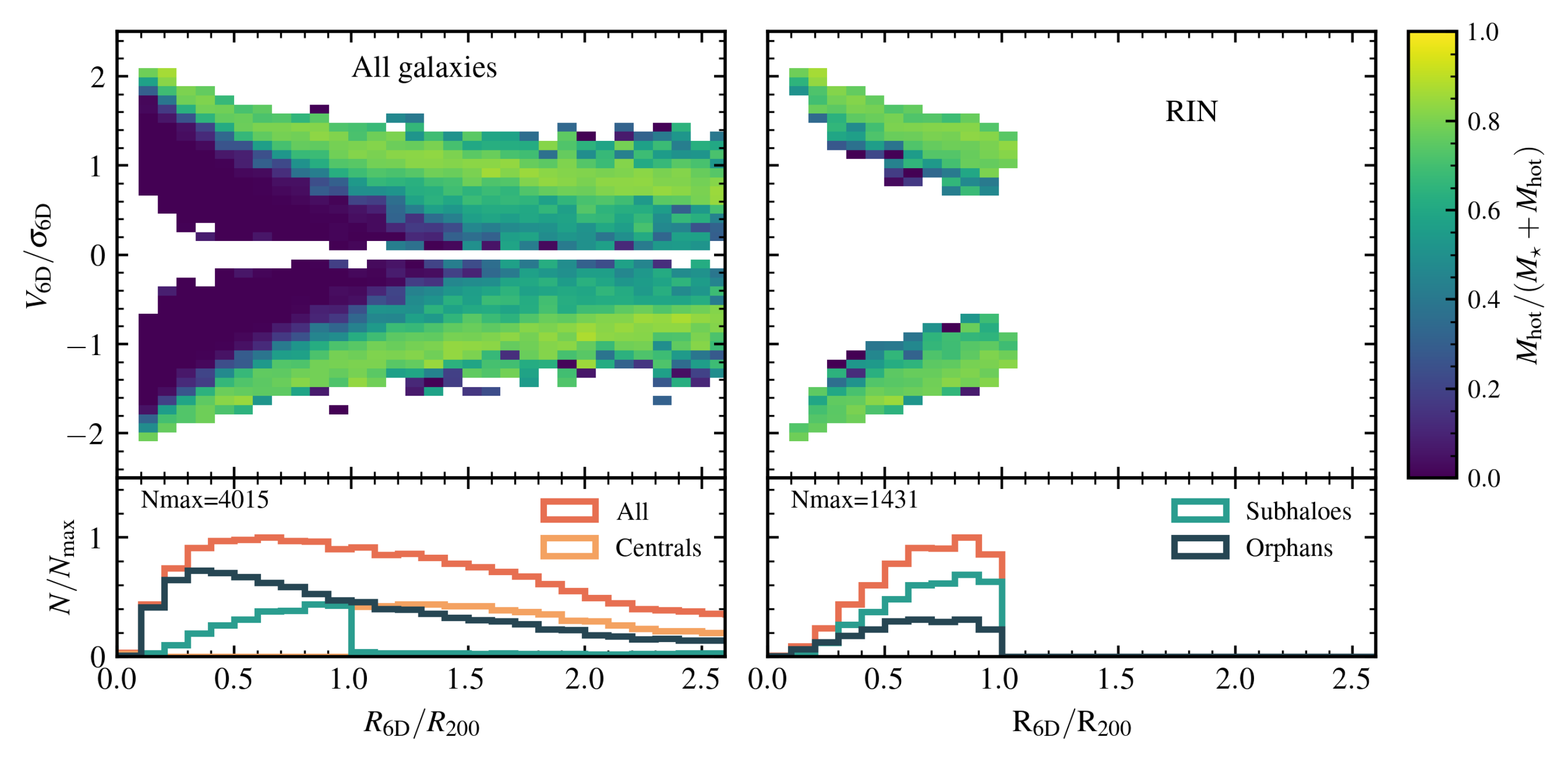}
    \includegraphics[width=0.8\textwidth]{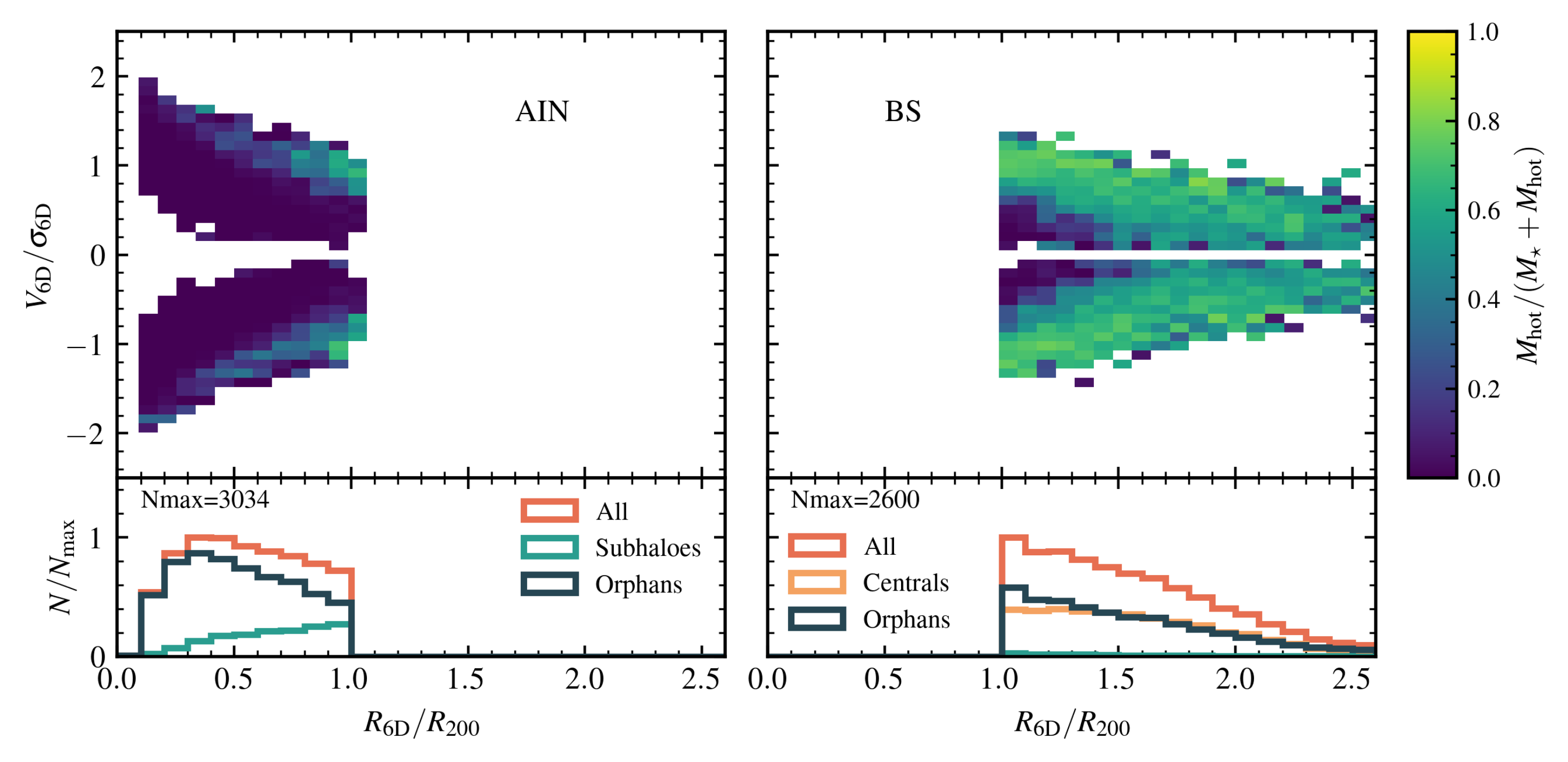}
    \caption{
    Correlation between phase-space position and hot gas content, represented by the hot gas fraction $f_{\rm hot}= M_{\rm hot}/(M_\star+M_{\rm hot})$, of all galaxies in and around clusters (upper left), RIN (upper right), AIN (lower left) and BS (lower right).
    Histograms at the bottom of each phase-space diagram show
    the composition of the corresponding population in terms of central and satellite galaxies; the histograms show the radial distribution of all galaxies of a given class in the diagram (red line), 
    and the contribution of
    centrals (orange line), subhaloes (satellites with DM haloes, greenish line) and orphans (satellites without DM haloes, dark line).
    All histograms are normalized 
    with the maximum number of galaxies from all bins (where the red curve reaches $1$, given by $N_{\rm max}$). 
    }
    \label{fig:phase-space-hotgas-ratio}
\end{figure*}

As we have mentioned, the vast majority of BS have only one pericentric passage, and they spend less than $\sim 2$ Gyr inside the cluster. The main difference 
between RIN and BS
is that only around half 
of RIN experience a pericentric passage, while all of BS have at least one pericentric passage (by definition), with the peak of the pericentric distance distribution being at $r_{\rm per}\sim 0.3 R_{200}$, well inside the cluster \citep[and in consistency with][]{Wetzel_2014}. 
We highlight that for the population of
BS that are passive at $z=0$, $\sim 85$ per cent of them quenches their SF after their first pericentric passage, in consistency with \citet{oman21}; from those, it is noteworthy that $\sim 65$ per cent of 
BS quenches their SF after leaving the cluster. 
The remaining $\sim 15$ per cent of BS that ceases to form stars before reaching their pericentre are
probably affected by pre-processing; 
this aspect deserves further investigation and is beyond the scope of this work.
We also mention that the sharp increase of $f_{\rm q}$ that all populations show at infall for masses higher than $\sim 10^{10.4}$ is likely due to the interplay between the SN feedback and AGN feedback around this characteristic mass \citep{Cora_2018}.

\subsection{Gas content and phase-space diagram}
\label{subsec:gascontent}

The connection between the gas content of galaxies and their location in the phase-space diagram provides useful information to understand the impact of environmental effects on the SF history of galaxies.
We build the phase-space diagram in $6\rm D$ based on the approach presented in \citet{arthur19} and \citet{mostoghiu21}, using $3\rm D$ positions $\rm (x,y,z)$ and velocities $\rm (v_x, v_y, v_z)$, and we correlate the hot gas content of galaxies with phase-space position.
This is presented in Fig.~\ref{fig:phase-space-hotgas-ratio} for all galaxies in and around clusters (upper left panel), 
RIN (upper right panel), AIN (lower left panel) and BS (lower right panel).
We estimate the normalized distance relative to the cluster centre $R_{\rm 6D}=\sqrt{\rm {(x-x_c})^2+(y-y_c)^2 + (z-z_c)^2 }/R_{200}$, where $ {\rm (x_c,y_c,z_c)}$ are the coordinates of the galaxy located at the centre of the cluster. 
The normalized velocity relative to the central galaxy is computed as $V_{\rm 6D} = {\rm sgn}(\mathbf{r\cdot v}) \sqrt{ {\rm (v_x-v_{ x,c})^2 + (v_y-v_{y,c})^2 + (v_z-v_{z,c})^2}}/\sigma_{\rm 6D}$, where $\mathbf{r}$ and $\mathbf{v}$ are position and velocity vectors relative to the central galaxy, ${\rm (v_{x,c}, v_{y,c}, v_{z,c})}$ represents the velocity components of the cluster central galaxy, and $\sigma_{\rm 6D}$ is the velocity dispersion of the cluster, computed as the 
mean square root of the relative velocity of all satellite galaxies (both 
subhaloes and orphans) with $ M_\star \geq 10^9 \, {\rm M_\odot}$ located inside $1.5 \, R_{200}$ of the cluster halo.
The hot gas content is characterized by the hot gas fraction $f_{\rm hot}= M_{\rm hot}/(M_\star+M_{\rm hot})$.

We find that the hot gas content of galaxies located in and around galaxy clusters is associated to their phase-space position, specially for $r \lesssim 1.5 \, R_{200}$ (upper left panel of Fig.~\ref{fig:phase-space-hotgas-ratio}).
It can be appreciated that galaxies located in the outskirts of the clusters tend to have high proportion of hot gas, with $f_{\rm hot}\gtrsim 0.6$. Moreover, galaxies tend to keep their hot gas as they enter the cluster, even if they reach the inner regions at high velocities ($R_{\rm 6D}\sim 0.3 \, R_{200}$ and $V_{\rm 6D}\sim 2\, \sigma_{\rm 6D}$). These galaxies are mainly RIN
and are the satellites within the clusters that at $z=0$ have higher amounts of hot gas (as can be appreciated from the upper right panel). These galaxies still have high velocities relative to the central galaxy, and $\sim 50$ per cent of them have not experienced a pericentric passage. 
It is noteworthy
that these galaxies can retain a relatively high amount of hot gas after infall, even for those that are approaching the cluster centre at high velocities, where environmental effects are stronger (the branch of the distribution with negative relative velocity), and also those that have experienced their first pericentric passage (the branch with positive relative velocity). This fact probes that $2\,{\rm Gyr}$ result not enough to efficiently remove the hot gas halo by environmental effects.

Conversely, AIN ($t_{\rm lbk,infall}> 2$ Gyr) tend to distribute closer to the cluster centre and have lower velocities than RIN (lower left panel), mainly due to dynamical friction acting on galaxy orbits over large periods of time. This population is basically depleted of hot gas ($f_{\rm hot}\lesssim 0.2$). Environmental effects as RPS accumulate over time, and the amount of hot gas decreases as galaxies orbit inside the cluster. This translates in low gas cooling rates with the consequent reduction of SF. Thus, AIN account for the majority of the passive population inside the cluster ($\sim 85$ per cent), as we showed in the previous section (Fig.~\ref{fig:passive-radial}).

Galaxies located outside \rdos~can be either BS or neighbouring galaxies.
BS also tends to keep a moderate proportion of hot gas ($0.4<f_{\rm hot}<0.6$), which does not correlate strongly with phase-space position (lower right panel of Fig.\ref{fig:phase-space-hotgas-ratio}). However, a high proportion of orphan galaxies that near \rdos~are still gravitationally bound to the clusters and have low velocities and low proportion of hot gas (this is presented more clearly in Fig. 2 of the Supplementary material).
It is striking that, in a large fraction of BS, a hot gas halo is still present in a galaxy that has entered the hostile environment of a massive galaxy cluster.
The causes of this are explored more deeply in Sec.~\ref{subsec:gasremoval}.
This population inhabits the outskirts of clusters along with neighbouring galaxies that have never been closer to the cluster centre than the radius \rdos~of the clusters,
and which have higher amounts of hot gas than BS (not shown here).

On the one hand, 
we can identify neighbouring galaxies with those having high proportion of hot gas that are infalling for the first time, at $ R_{\rm 6D} > R_{200}$ and $ V_{\rm 6D}<-\sigma_{\rm 6D}$. On the other hand, neighbouring galaxies with positive velocities have similar hot gas content, but are receding away from the cluster ($V_{\rm 6D}<\sigma_{\rm 6D}$).
Thus, they are characterized by higher relative velocities 
which can be inferred by the difference between the upper left and bottom right panels.

It is worth noting that the cold gas fraction, computed as $f_{\rm ColdGas} = M_{\rm ColdGas}/(M_\star+M_{\rm ColdGas})$, follows approximately the same distribution in the PSD that hot gas, but with values of $f_{\rm ColdGas}\lesssim 0.4$ (not shown here). In our model, 
the cold gas results more difficult to be removed by RPS, as the hot gas acts as a shield of the former, and $90$ per cent of the 
hot has has to be removed before RP is able to act on the cold gas reservoir (see Sec.~\ref{sec:model}). 

Our model produces substantial differences with respect to the hydrodynamical simulation \textsc{gadget-x} \citep{springel2005} run in the context of \textsc{The Three Hundred} project \citep{cui2016, arthur19,mostoghiu21}. In particular, \citet{arthur19} show that the gas fraction decreases abruptly as galaxies approach the cluster, and is basically depleted of gas once it crosses a `shock radius' of $\sim 1.5 \, R_{200}$. \citet{mostoghiu21} show that the gas content of haloes is higher than in subhaloes (likely due to pre-processing over the latter) and that a small fraction of them can retain their gas after $\sim 4-5\,{\rm Gyr}$ after entering the infall region (defined as the region inside $4\, R_{200}$ from the cluster centre, in their work). In addition, they 
also find an accretion shock between $\sim 1.5-2 \, R_{200}$, beyond which galaxies completely lose their gas with respect to the moment of infall (see their figs. 6 and 7). 
These differences can be explained by taking into account that, in the \sag~model, there is a spatial limit to the environmental effects exerted by the cluster, which is not present in hydrodynamical simulations. That is,
\sag~includes environmental effects for satellite galaxies identified by the halo finder, which are mainly located within \rdos~of the haloes (see upper part of Table~\ref{Table_Percentage} for the case of the galaxy clusters in our sample).
Satellites of neighbouring centrals are also affected by the environmental effects exerted by their haloes, which are milder consistent with their lower mass. However, there is not a continuous transition of such effects between the clusters and the neighbouring haloes in their outskirts.

The histograms under each diagram of Fig.~\ref{fig:phase-space-hotgas-ratio} show how central galaxies, subhaloes 
and orphans compose the different populations of galaxies (the histogram of the upper left panel
is equivalent to
Fig.~\ref{fig:histo-r-Type1Type2}). 
In the upper right panel,
we see that RIN are spatially distributed towards \rdos, and are dominated by satellites with DM haloes (subhaloes)
for $r \gtrsim 0.5\, R_{200}$ as
TS produced by the deep potential well of the cluster halo cannot remove efficiently the DM halo, and they are still detected over the resolution limit of the simulation;
in the inner regions, the distribution of orphans is similar to that of subhaloes.
AIN tend to distribute near the central galaxy since the relative velocity of AIN is reduced by the action of dynamical friction as a result of the longer period of time spent within the cluster.
Orphan galaxies outnumber subhaloes at all distances inside the cluster for this population, likely due to a combination of dynamical friction acting on longer timescales and TS that removes DM particles of the halo, which can no longer be detected by the halo finder.
In the outskirts of the cluster, the population of BS is divided in almost equal parts between centrals and orphan galaxies (see Table~\ref{Table:Percentage2}), whose distributions decline monotonically with increasing distance; subhaloes represent a marginal population. 
Half of those orphan BS are still associated to the cluster (see upper part of Table~\ref{Table_Percentage}), as the orbital evolution model used to integrate their orbits allows this type of galaxies to orbit further than \rdos, and are likely to become cluster satellites later on \citep{gill05, Wetzel_2014,muriel2014, Diemer_2021}. 

In summary, Fig.~\ref{fig:phase-space-hotgas-ratio} 
shows that, on average, galaxies that have lost their hot gas halo are basically located inside the cluster. However, not only location is important in gas stripping, but the complete dynamical history and the time spent inside the cluster is fundamental for the effective action of RP. This is further analysed in the next section, as well as in the Supplementary material.

\begin{figure*}
    \centering
    \includegraphics[width=\textwidth]{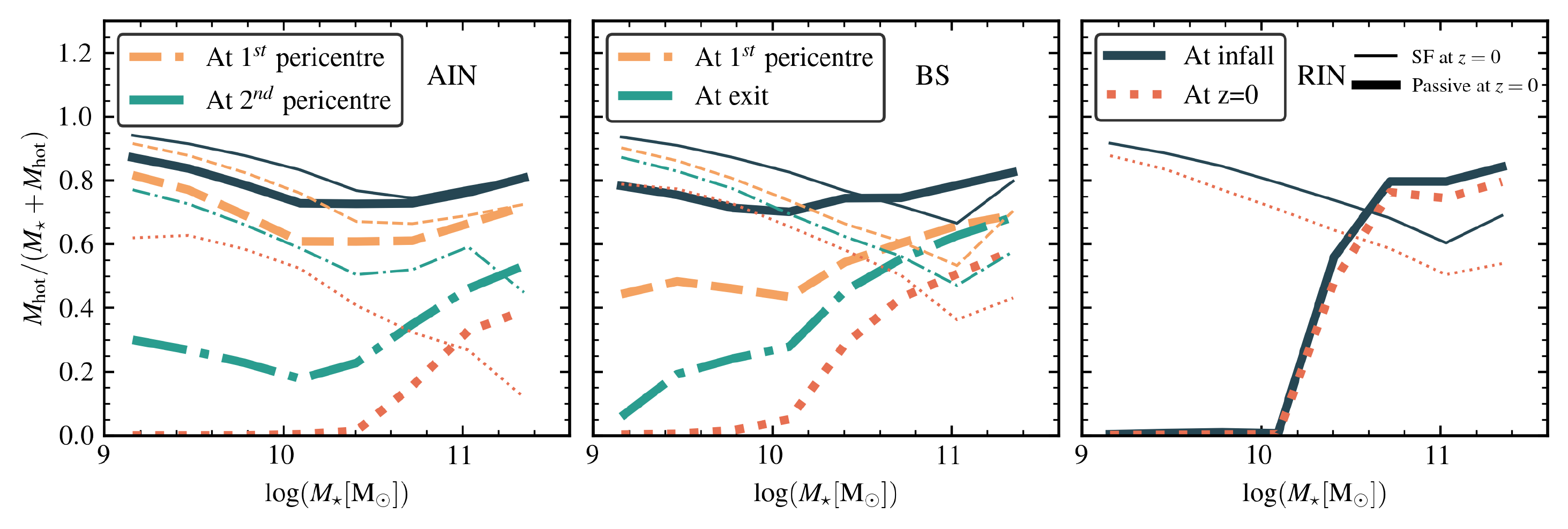}
    \caption{
    Median values of the fraction of hot gas, $f_{\rm hot} = M_{\rm hot\, gas}/(M_\star+M_{\rm hot\,gas})$, as a function of stellar mass at $z=0$ for AIN (left panel), BS (middle panel) and RIN (right panel) at different moments of their orbital evolution: at infall time (solid dark line) and at $z=0$ (red dotted line) for all galaxy populations; at the moments of first and second pericentric passages for AIN (orange dashed and green dashed-dotted lines, respectively); at the moments of first pericentric passage and exit from the cluster for BS (orange dashed and green dashed-dotted lines, respectively).
    Each population is separated in those that are passive at $z=0$ (thick lines) and those that are star-forming at $z=0$ (thin lines).
    }
    \label{fig:fhot-mstar_AIN-RIN-BS}
\end{figure*}

\begin{figure*}
    \centering
    \includegraphics[width=\textwidth]{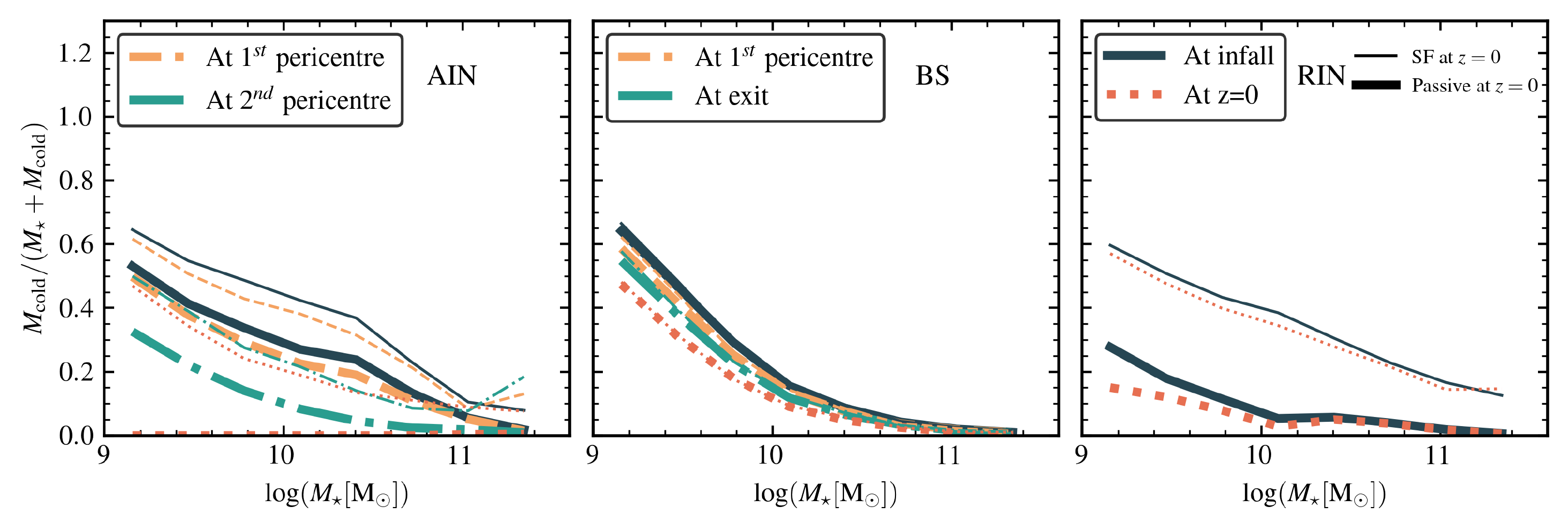}
    \caption{
    Median values of the fraction of cold gas, $f_{\rm  cold} = M_{\rm cold\,gas}/(M_\star)$, 
    as a function of stellar mass at $z=0$ for AIN (left panel), BS (middle panel) and RIN (right panel) at different moments of their 
    orbital evolution: at infall time (solid dark line) and at $z=0$ (red dotted line) for all galaxy populations; at the moments of first and second pericentric passages for AIN (orange dashed and green dashed-dotted lines, respectively); at the moments of first pericentric passage and exit from the cluster for BS (orange dashed and green dashed-dotted lines, respectively).
    Each population is separated in those that are passive at $z=0$ (thick lines) and those that are star-forming at $z=0$ (thin lines).
    Low-mass galaxies tend to keep a higher proportion of cold gas than high-mass galaxies, even if they are quenched. 
    }
    \label{fig:fcold-mstar-AIN-RIN-BS}
\end{figure*}


\subsection{Gas removal}
\label{subsec:gasremoval}

In order to explore the impact of RPS on the hot gas content, we analyse the evolution of the fraction of hot gas, $f_{\rm hot} =  M_{\rm hot}/(M_\star+M_{\rm hot})$, in a similar manner as in Fig.~\ref{fig:fq-mstar_AIN-RIN-BS}. 
Fig.~\ref{fig:fhot-mstar_AIN-RIN-BS} shows how $f_{\rm  hot}$ evolves in passive and star-forming (at $z=0$) AIN, BS and RIN, respectively, since the moment of infall until $z=0$, for different stellar masses estimated at $z=0$. 
We find a clear consistency between the decline of the hot gas content of passive low-mass galaxies and the increase of the fraction of quenched galaxies shown in Fig.~\ref{fig:fq-mstar_AIN-RIN-BS}.

As can be seen from the left panel of Fig.~\ref{fig:fhot-mstar_AIN-RIN-BS}, at infall, passive AIN present 
a mild dependence
of $f_{\rm  hot}$ on stellar mass, with a slight decline from $f_{\rm hot}\sim 0.9$ at $M_\star \sim 10^{9} {\rm M_\odot}$ to $f_{\rm hot}\sim 0.75$ at $M_\star \sim 10^{10.8}$, and a small increase of the trend for higher stellar masses (thick dark solid line); this trend is still present by the time galaxies reach the first pericentre of their orbit, although with slightly smaller values of $f_{\rm  hot}$ (thick orange dashed line). 
After this stage, the hot gas content is strongly reduced by the time they reach their second pericentre, specially for low-mass galaxies where $f_{\rm hot}\sim 0.25$.
At ${z=0}$, passive low-mass AIN are depleted of hot gas, while high-mass AIN still preserve a moderate amount of it.
The current low-mass star-forming population of AIN is somehow still able to keep relatively high amounts of hot gas, as they present a more gradual evolution of their fraction of hot gas (thin lines), and by $z=0$ they have $f_{\rm hot}\sim 0.6$ (thin red dotted line).

The evolution of the hot gas fraction of BS (middle panel of Fig.~\ref{fig:fhot-mstar_AIN-RIN-BS}) resembles that of AIN. For low-mass passive BS, $f_{\rm hot}$ evolves dramatically from $\sim 0.8$ at infall to $0$ at ${z=0}$ (thick lines), although it seems to have a more continuous decline prior and after the first pericentric passage; 
note that, for BS, we consider the moment in which the galaxy leaves the cluster instead of the second pericentric passage as in the case of AIN.
High-mass passive BS present little evolution of $f_{\rm hot}$, from $\sim 0.8$ to $\sim 0.5$. 
Current star-forming BS present little evolution of their hot gas content regardless of their $z=0$ stellar mass (thin lines), and present an anti-correlation with stellar mass from $\sim 0.8-0.9$ for low-mass BS to $\sim 0.4-0.7$ for high-mass BS.

On the contrary, passive and star-forming RIN (right panel of Fig.~\ref{fig:fhot-mstar_AIN-RIN-BS}) present little evolution of the hot gas content (an average decrease of $\sim 5-10$~per cent for star-forming RIN, regardless of the stellar mass content), in consistency with the almost null evolution of the fraction of quenched satellites in the last $2\,{\rm Gyr}$. We remark that current low-mass passive RIN arrive at the cluster with null content of hot gas, meaning that pre-processing effects 
have strong impact on
this subsample of galaxies (see Section ~\ref{sec:transformation}); 
in contrast, star-forming RIN have high proportions of hot gas, from $f_{\rm hot}\sim 0.9$ for low-mass galaxies to $f_{\rm hot}\sim 0.6$ for high-mass ones (thin lines).

Regarding the cold gas content of galaxies,  Fig.~\ref{fig:fcold-mstar-AIN-RIN-BS} shows the dependence of the cold gas fraction, $f_{\rm cold}=M_{\rm cold}/(M_\star + M_{\rm cold})$, on $z=0$ stellar mass and how it evolves through different timestamps as in the previous analysis, 
separating AIN, BS and RIN into passive and star-forming galaxies at $z=0$.
For the star-forming population identified in all samples, we notice an overall trend of $f_{\rm cold}$ decreasing as stellar mass increases, consistent with the general observed trend used to calibrate the model (\citealt{Boselli14}, see fig. 3 of \citealt{Cora_2018}). 
The cold gas left 
in high-mass galaxies of both passive and star-forming populations (the latter represents a small fraction, see Fig.~\ref{fig:fq-mstar_AIN-RIN-BS}) is lower because AGN feedback prevents the replenishment of the cold gas reservoir and it has been practically completely consumed by SF by the time of infall (see \citealt{cora2019} for a more detailed discussion). 
Since the evolution of the cold gas fraction becomes more pronounced as stellar mass decreases, the following analysis is focused on low-mass galaxies.  

For passive low-mass AIN (left panel of Fig.~\ref{fig:fcold-mstar-AIN-RIN-BS}, thick lines), $f_{\rm cold}$ has a negligible decrease between infall and first pericentric passage, but decreases from $\sim 0.5$ to $\sim 0.3$ between the first and second pericentres, reaching null values
at the present (orange dashed, green dashed-dotted and dotted red lines, respectively).
These galaxies become depleted of cold gas because their hot gas halo is reduced substantially ($f_{\rm hot}<0.10$).
Thus, on the one hand, there is no more gas cooling that fuels the cold gas reservoir and, on the other, the cold gas disc can be stripped by RP since it is no longer shielded by the hot gas halo. Hence, the combined effect of RPS, SF and SN feedback contributes to the cold gas depletion.
Moreover, since our quiescent
SF scheme adopts a mass threshold for the cold gas below which no stars form (see Sec.~\ref{subsec:SF}),
it leaves a residual amount of cold gas
that can be completely removed by SN feedback and/or RPS.

The time spent by a satellite inside a cluster, being affected by the action of RP, is key in the evolution of its gas content.
The cold gas fraction of RIN (right panel of Fig.~\ref{fig:fcold-mstar-AIN-RIN-BS}) remains basically unchanged between infall and ${z=0}$, for both the passive (thick lines) and the star-forming (thin lines) populations.
As $f_{\rm hot}$ 
remains high for the star-forming population, RP is not able to act directly on the cold gas phase, which keeps a fraction $f_{\rm cold} \sim 0.6$ that likely results from the balance between gas cooling, SF and SN feedback. Instead, the lower fractions of cold gas that characterise the passive low-mass RIN ($f_{\rm cold} \sim 0.2-0.3$)
are linked to their mean null fraction of hot gas, that is, the cold gas reservoir of these galaxies are not longer fueled by gas cooling and it is simply reduced by SF and/or SN feedback. We have verified that the fraction of cold gas stripped by RP is null in these galaxies, so RPS is discarded as a mechanism that contributes to reduce the cold gas disc in this galaxy population. 

The cold gas fractions of BS (middle panel of Fig.~\ref{fig:fcold-mstar-AIN-RIN-BS}) present a particular aspect that differs from the patterns shown by AIN and RIN. While passive galaxies of AIN and RIN have lower fractions of cold gas than their star-forming galaxies, for a given stellar mass and stage along their orbital evolution, these values are almost the same for both the passive and star-forming populations of BS.
In other words, the cold gas fractions of AIN, RIN and BS have similar trends and values for the star-forming populations, while they differ substantially for the passive ones.
This particular behaviour is explained by focusing on the distributions of lookback infall times (Fig.~\ref{fig:histo-time-inside}) and the time spent by the galaxies inside clusters.
On the one hand, the peak of the lookback infall time distribution of BS lies between those of AIN and RIN, with a mean value of $4.1 \, {\rm Gyr}$ \citep[in consistency with the results found by][]{Wetzel_2014}, but this distribution comprises values corresponding to RIN and AIN, being dominated by the later.
Thus, the fractions of cold gas at infall of BS resemble more those of AIN, because they have less time to be affected by pre-proccesing.
On the other hand,
BS have spent relatively short times  within the cluster (mean value $1.52 \, {\rm Gyr}$), which are of the order of the mean lookback infall times of RIN (mean value $0.93 \, {\rm Gyr}$). This explains the slight decrease of $f_{\rm cold}$ with time for a given stellar mass displayed by both RIN and BS. This mild evolution prevents BS from reaching at $z=0$ the almost null fractions of cold gas typical of AIN. 
Although AIN and BS show pretty similar dependence of the fraction of hot gas on stellar mass, with and abrupt reduction after the first pericentric passage, and median null values achieved at $z=0$ for low-mass galaxies (see Fig.~\ref{fig:fhot-mstar_AIN-RIN-BS}),
passive BS can keep similar amounts of cold gas to those of star-forming BS (a situation similar to that discussed for low-mass passive RIN).

This analysis allows us to conclude that, as low-mass galaxies lose their hot gas as a result of RPS, the gas cycle is interrupted and the cold gas reservoir available to form stars is consumed by SF and/or ejected by SN, and eventually removed by RPS in the case of AIN. 
This probes that
the absence of hot gas in low-mass galaxies is a necessary condition for SF quenching in haloes of $M_{200}\gtrsim 10^{15} \, {\rm M_\odot}$.
At first glance,
this conclusion contrasts with the results obtained by \citet{cora2019}, who find that a sample of low-mass satellites 
still preserves a moderate proportion of hot gas at the moment of quenching (see their fig. 8), and their SF is suppressed due to a strong reduction in the gas cooling rate of the hot gas. However, their results where obtained for a sample of satellites located within a mixture environments characterized by a range of masses below those typical of our current sample of massive clusters ($10^{12.5}<M_{200}\,[ {\rm M_\odot}]<10^{15}$).
All in all, the key point for SF quenching is the reduction or suppression of gas cooling that feeds the cold gas reservoir, and this takes place in low-mass galaxies by the removal of the hot gas halo by RPS. It is worth mentioning that there are low-mass galaxies (like most of BS) that still preserve cold gas with fractions of $f_{\rm cold} \gtrsim 0.3$ when they quench. 
This means that our conclusions rely not only on the RPS model but also on the SF model, i.e., the fact that the cold gas content must surpass a minimum threshold in order to form stars allows a galaxy to be quiescent 
despite of its non-negligible content of cold gas.

\subsection{Impact of pericentric passages}
\label{subsec:pericentre}

The pericentric passage is a particular moment of the satellite's evolution, and is considered as a milestone \citep{Lotz_2019, DiCintio_2021, oman21,benavides2021, moreno2022, smith2022}. The minimum distance to the cluster centre that a galaxy can reach is generally close to the peak in the density of the ICM through which it is moving. Depending on the relative velocity, the RP exerted over the galaxy can reach a maximum value during the pericentric passage.
To identify where and when a pericentric passage occurs, we follow the evolution of the distance 
of each galaxy to the cluster centre and calculate the local minima of the distance that are located inside \rdos~(see Fig.~\ref{fig:distance-evolution}).

In Figs.~\ref{fig:fq-mstar_AIN-RIN-BS} and \ref{fig:fhot-mstar_AIN-RIN-BS}, we show that both the fraction of quenched galaxies and 
hot gas content change dramatically after the first pericentric passage, specially for AIN and passive BS. However, the almost null evolution of RIN might indicate that some time ($\gtrsim 1\, {\rm Gyr}$) must elapse after the passage before there is an impact on galaxy properties. 

\begin{figure}
    \centering
    \includegraphics[width=\columnwidth]{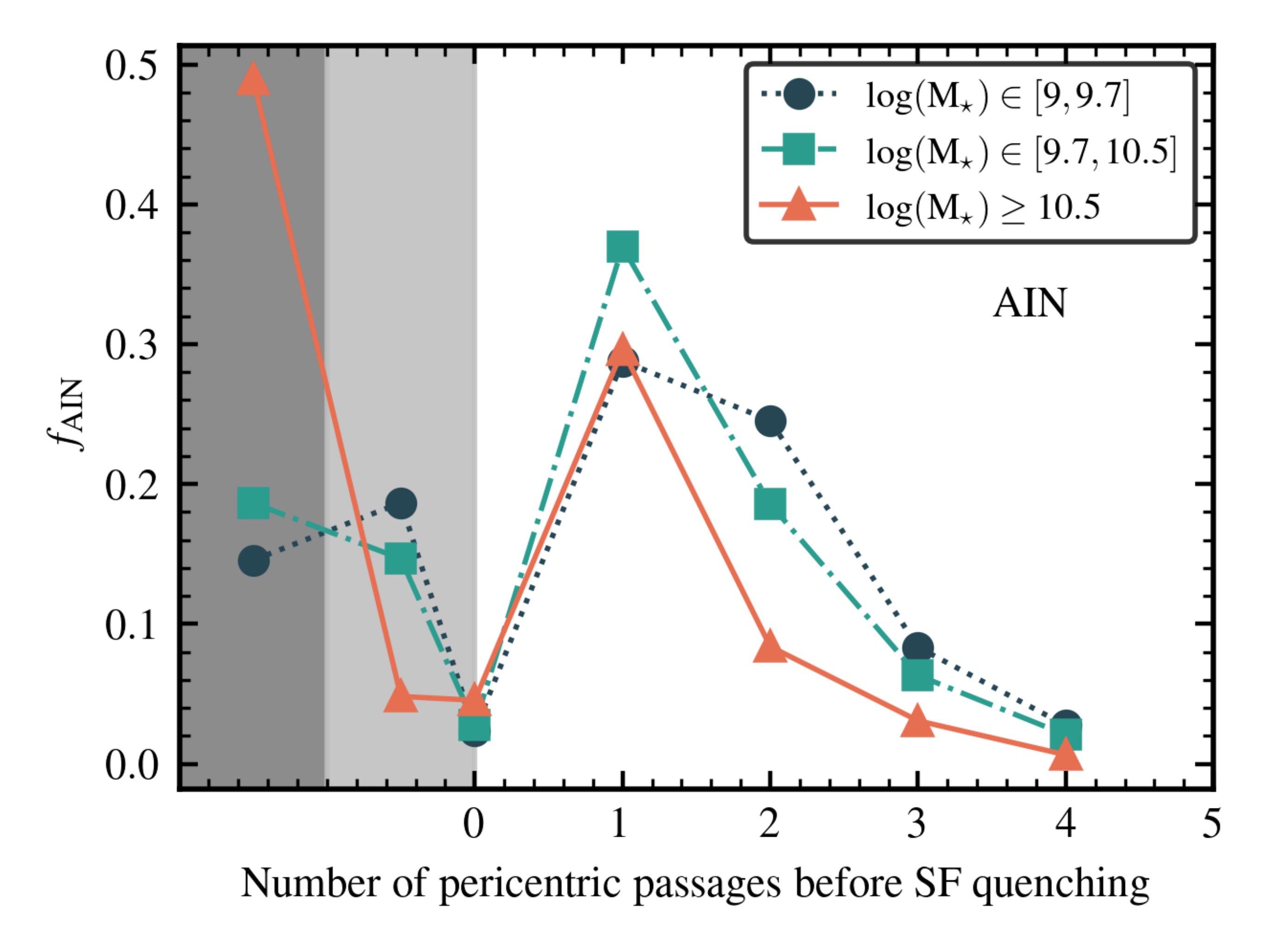}
    \caption{Fraction of AIN that are star-forming at infall and quenched at ${z=0}$, as a function of the number of pericentric passages that experience before SF quenching occurs. 
    They are indicated by different symbols (connected by thin lines) corresponding to different stellar mass ranges, as detailed in the legend.
    Regardless of stellar mass, the majority of AIN quenches after one pericentric passage. A non-negligible fraction of low-mass AIN needs a second pericentre passage before quenching occurs. 
    For completeness, the value of $f_{\rm AIN}$ located over the dark-gray area represents the fraction of galaxies of a given mass that quenches before crossing \rdos, while the value located over the light-gray area represents those AIN that are star-forming both at infall and ${z=0}$ (they are not quenched yet).}
    \label{fig:perpas}
\end{figure}

In Fig.~\ref{fig:perpas}, we show the fraction of AIN, $f_{\rm AIN}$, that are 
star-forming at infall and quenched at ${z=0}$ as a function of the number of pericentric passages they experience before they quench,  that is, we identify the snapshot when an AIN becomes passive, and count how many pericentric passages they experience inside the cluster before that snapshot. 
On the one hand, a negligible fraction of AIN quenches before reaching the pericentre (less than $\sim 5$ per cent, regardless of stellar mass). On the other hand, the majority of AIN quenches between first and second pericentric passages ($f_{\rm AIN}\sim 0.3-0.35$), indicating the importance of this transition. However, a non-negligible fraction of AIN quenches between the second and third passages, particularly for the lower mass bins ($f_{\rm AIN}\sim 0.2$, while for the high-mass bin $f_{\rm AIN}\sim 0.1$). It has been established that the quenching timescale decreases with
increasing stellar mass due to the fact that high-mass galaxies experience not only environmental effects but also secular effects \citep{peng2010,Wetzel_2014,hahn2017, cora2019}. This means that high-mass galaxies quench earlier (on average) than low-mass ones,
consistent with the downsizing scenario \citep{noeske2007,cattaneo2008}.
It is worth noting that there are AIN quenched prior to infall (before crossing \rdos), as it is evident from the left panel of Fig.~\ref{fig:fq-mstar_AIN-RIN-BS} 
which shows
quenched fractions of the order of $\sim 0.5$ and $\sim 0.2$ for high-mass and low-mass AIN, respectively; this is denoted by the fractions of $f_{\rm AIN}$ on the dark-gray region of Fig.~\ref{fig:perpas} to which negative numbers of pericentric passages are assigned. 
The symbols on the light-gray region, also characterized by negative values, represent those AIN that have not quenched yet, i.e. they are star-forming both at infall and ${z=0}$.

\begin{figure}
    \centering
    \includegraphics[width=\columnwidth]{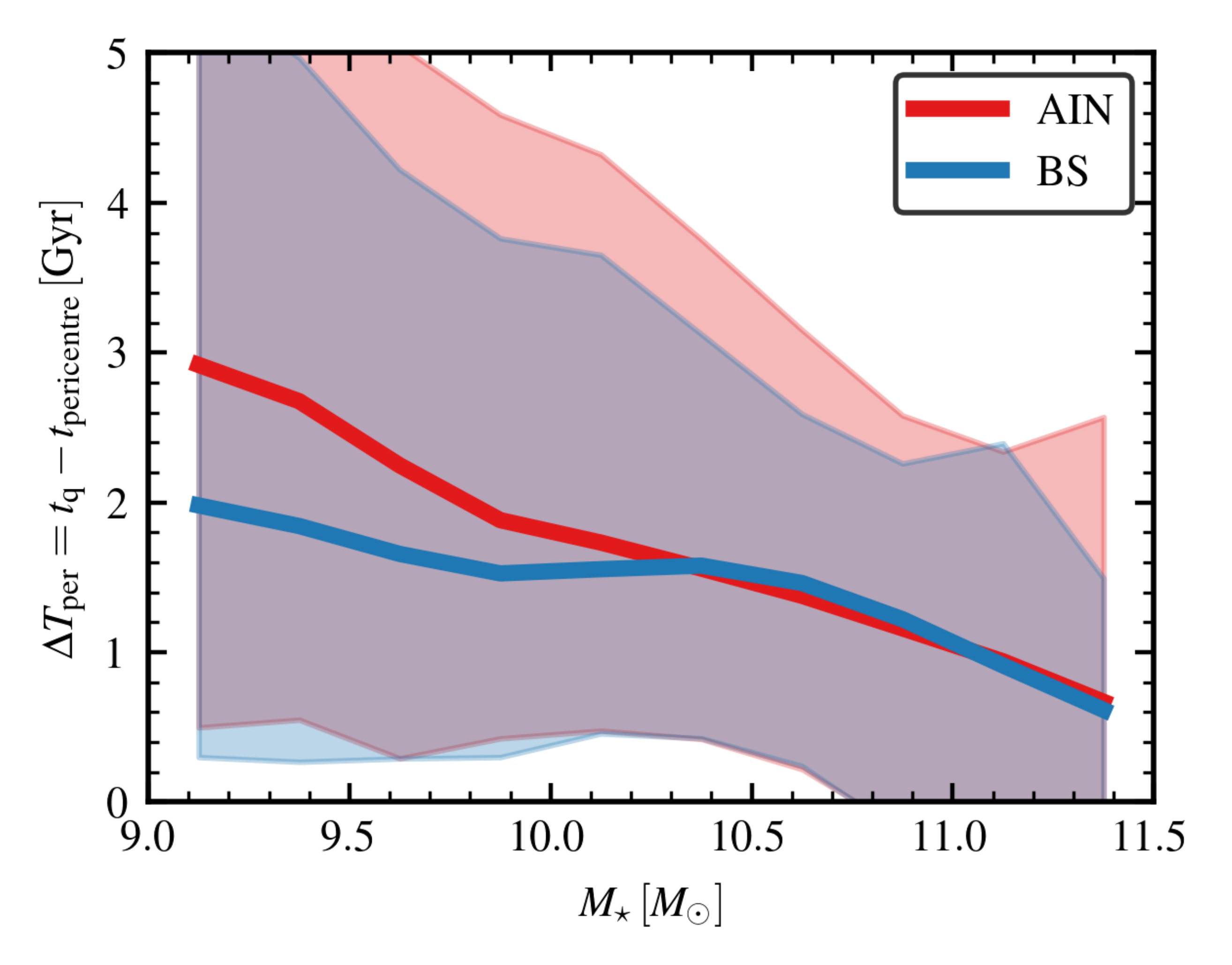}
    \caption{Time delay between first pericentric passage ($t_{\rm pericentre}$ and the moment of SF quenching ($t_{\rm q}$), as a function of stellar mass, for AIN and BS that cross \rdos~as star-forming galaxies, and result passive by ${z=0}$. Solid lines represent the median of the distributions (as indicated in the legend), while shaded regions denote $10-90$ percentiles. 
    }    \label{fig:tdelay}
\end{figure}

In Fig.~\ref{fig:tdelay}, we explore the 
period of time elapsed
between the first pericentric passage and the moment of SF quenching, for both AIN and BS. 
This analysis has no sense for RIN since their quenched status at $z=0$ is already established at infall.
We refer to this timescale as a delay time, $\Delta T_{\rm delay}$. 
We find that the distribution of $\Delta T_{\rm delay}$ presents median values that decrease with increasing stellar mass (solid lines), and a broad scatter denoted by the $10-90$ percentiles (shaded regions). While for higher masses ($M_\star \gtrsim 10^{10.5}\, {\rm M_\odot}$) both AIN and BS quench between $0.6-1.6\,{\rm Gyr}$ 
after the first pericentric passage, low-mass BS tend to have $\Delta T_{\rm delay}\sim 2\,{\rm Gyr}$, between $\sim 0.4-1\,{\rm Gyr}$ lower than AIN. This difference is likely to arise because passive BS present a somewhat more abrupt evolution of their hot gas content while they orbit inside the cluster, giving place to an overall shorter quenching timescale. 
Curiously, the median values of
the distance at first pericentric passage is similar for passive BS and AIN ($r_{\rm per}/R_{200} \sim 0.34 $), and for star-forming BS and AIN ($r_{\rm per}/R_{200}\sim 0.4$); however, we find that passive BS can reach higher relative velocities along their orbits (the distribution of velocities peaks at $\sim 3100 \, {\rm km s^{-1}}$) than passive AIN (the distribution peak is at $\sim 2600 \, {\rm km s^{-1}}$). This is expected, as BS reach velocities high enough to be able to exit the cluster.

Our median values of $\Delta T_{\rm delay}$ are 
lower than in \citet{oman21}.
Using the Hydrangea hydrodynamic simulations \citep{schaye2015,bahe2017}, they find that $t_{\rm delay}\sim 4-5\,{\rm Gyr}$ with no clear dependence on stellar mass for clusters with $M_{200} > 10^{14} {\rm M_\odot}$. Moreover, they find that neutral hydrogen disappears before or around first pericentric passage, but 
SF
activity persists for $\sim 5\,{\rm Gyr}$ , when the satellite is around their second (or subsequent) pericentre.
On the contrary, using the Magneticum Pathfinder hydrodynamic simulation, \citet{Lotz_2019} find extremely short quenching timescales, as almost all satellites quench within $1\,{\rm Gyr}$ after infall to a 
cluster with
$M_{\rm vir} > 10^{14} {\rm M_\odot}$, by the time they are reaching their first pericentre. 

To summarize, we find that the pericentric passage 
leaves its imprint on the subsequent evolution of the SF activity of satellites and related properties,
although its influence is not extremely determinant. For instance, while most galaxies quenches after the first pericentric passage, an important fraction of low-mass galaxies quenches after their second pericentric passage; this is consistent with the high dispersion of the distribution of $t_{\rm delay}$ for low stellar masses. 
In the case of BS, all of them experience one passage through the cluster, but the combination of environmental effects inside the cluster with pre-processing that takes place before infall is what settles the condition for quenching to occur outside \rdos~after leaving the cluster. This is explored in the next section.

\section{Physical processes involved in galaxy transformation}
\label{sec:transformation}

\begin{figure*}
    \centering
    \includegraphics[width=\textwidth]{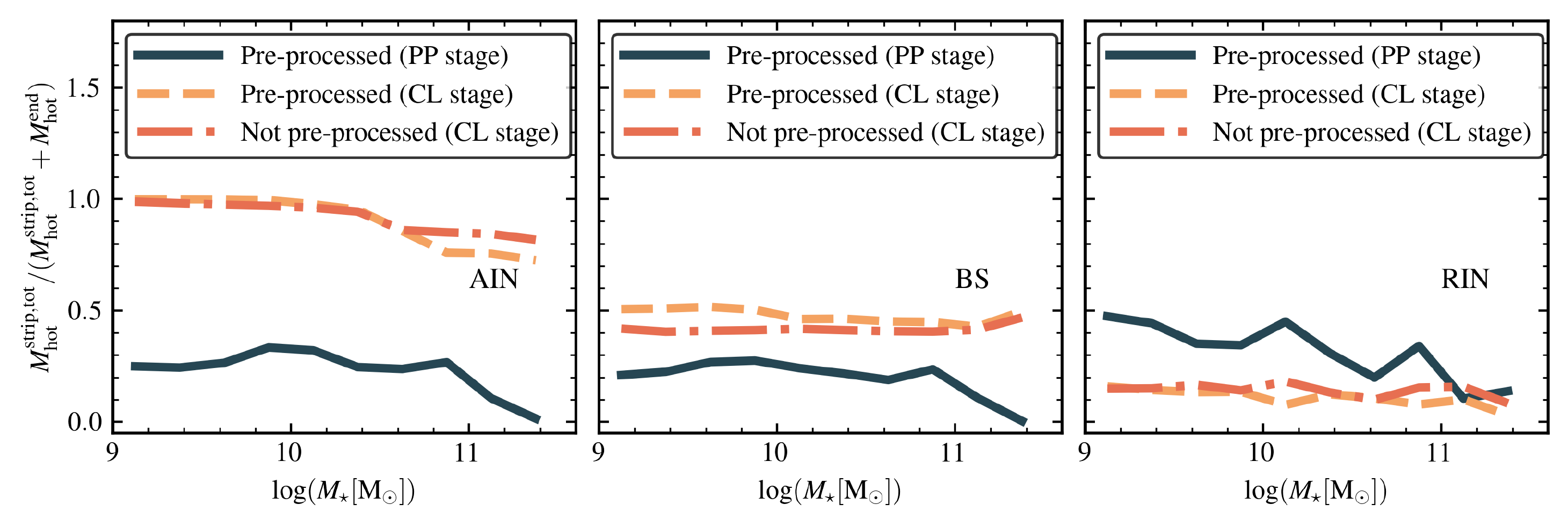}
    \caption{Median values of the fraction of hot gas removed by RPS for AIN, BS and RIN at different stages of their evolution. We separate them in those that have and have not experienced pre-processing, and compute the amount of hot gas removed in the pre-processing (PP) stage and in the cluster (CL) stage. We define $f_{\rm RPS} = M_{\rm hot}^{\rm strip,tot}/(M_{\rm hot}^{\rm strip,tot}+M_{\rm hot}^{\rm end})$, where $M_{\rm hot}^{\rm strip,tot}$ is the total mass of hot gas stripped during a given stage, and $M_{\rm hot}^{\rm end}$ is the hot gas available at the end of that stage (see text for details). 
    We calculate $f_{\rm RPS}$ for AIN (left panel), BS (middle panel) and RIN (right panel) separating the galaxies that experience pre-processing and are affected in both the PP stage (dark solid line) and once they enter the cluster (CL stage, orange dashed line), and those galaxies that only experience RPS inside the cluster (red dashed-dotted line).
    }
    \label{fig:gasstrip}
\end{figure*}

\begin{figure*}
    \centering
    \includegraphics[width=\textwidth]{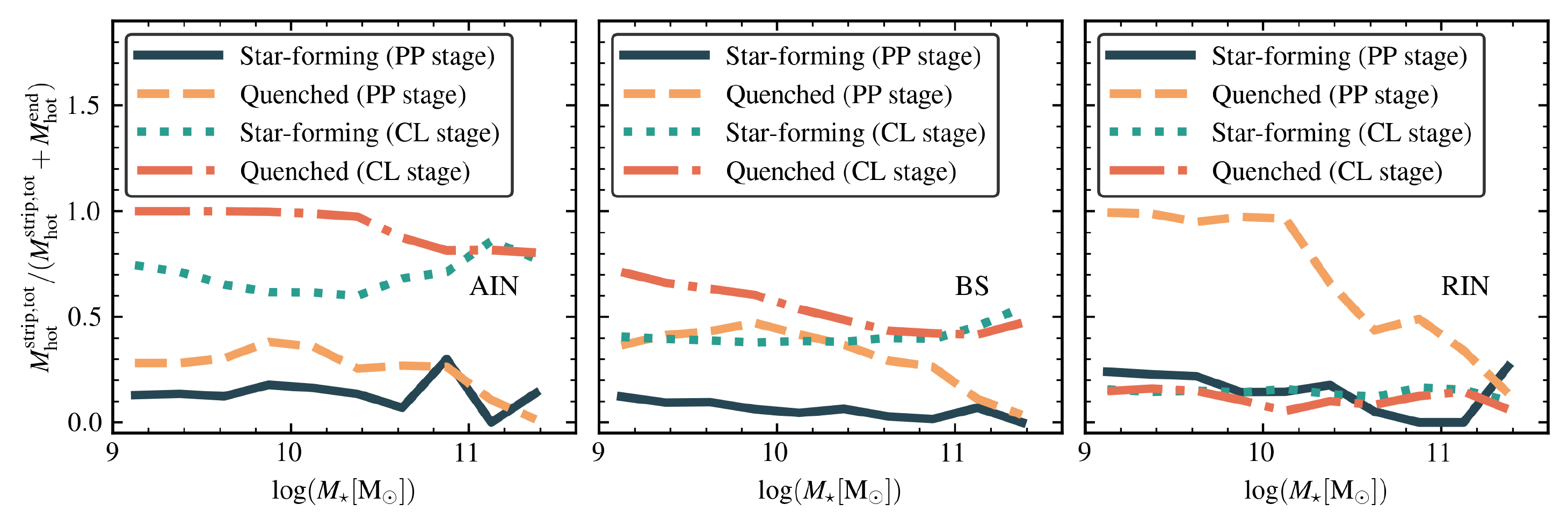}
    \caption{Median values of the fraction of hot gas removed by RPS for AIN, BS and RIN at different stages of their evolution.
    We calculate $f_{\rm RPS}$ 
    for AIN (left panel), BS (middle panel) and RIN (right panel), but separating them in star-forming and passive galaxies at $z=0$ and considering the RP experienced by them  during the PP and CL stages. Thus, we calculate $f_{\rm RPS}$
    for star-forming galaxies during the PP stage (dark solid line) and the CL stage (greenish dotted line), and for passive galaxies during the PP stage (orange dashed line) and the CL stage (red dashed-dotted line).}
    \label{fig:gasstrip2}
\end{figure*}

It is expected that galaxies that orbit closer to the cluster centre, or spend more time inside the cluster, would be more prone to lose gas due to RPS \citep{abadi1999,steinhauser2016,vega2022}. 
However, it is not well established the impact of physical processes that act on satellites of different haloes prior to infall into the cluster, usually known as pre-processing.
This processes can take place in group-like environments \citep[][and references therein]{rodriguez2022} and/or filament-like environments \citep[][and references therein]{kuchner2020}.
Simulations suggests that different host environments may led to different transformation pathways, timescales, and thus pre-processing efficiencies \citep{donnari2021b}. Though the controversy is far from solved, simulations \citep{bahe2017} and observations \citep{kuchner2017,bianconi2018} alike confirm that galaxies experience significant gravitational perturbations and hydrodynamical interactions beyond $2-3\, R_{200}$, long before reaching the cluster core.

In this work, we refer as pre-processing to
the effect of RPS over a satellite while it orbits inside a halo prior to infall to the main progenitor of the cluster halo. 
We define two stages in satellite evolution, namely the `pre-processing stage' (PP stage) and the `cluster stage' (CL stage), and use two different infall times to define the beginning of these stages: 1) the time of first infall to any halo different from the main progenitor of the cluster, for the PP stage, characterized by a lookback time, $t_{\rm lbk,infall}^{\rm first}$, and 2) the time 
of infall to the main progenitor of the cluster halo, for the CL stage, which
is the time of infall that we have been using in the previous analysis, and is characterized by a lookback time $t_{\rm lbk,infall}$, already defined. Note that $t_{\rm lbk,infall}$ is determined by the moment in which a galaxy crosses \rdos~of the main progenitor halo, while the estimation of $t_{\rm lbk,infall}^{\rm first}$ relies in the identification of the change of the galaxy condition, from central to satellite, made by the halo finder.
Thus, the period of time comprised
between $t_{\rm lbk,infall}^{\rm first}$ and $t_{\rm lbk,infall}$ defines
the pre-processing stage,
while the period of time elapsed from $t_{\rm lbk,infall}$
defines the cluster stage;
for AIN and RIN, the end of the cluster
stage is $z=0$, but for BS it is the moment each galaxy exits the cluster. 
With these definitions, 
we classify galaxies that do not experience pre-processing as those with $t_{\rm lbk,infall} = t_{\rm lbk,infall}^{\rm first}$; all these galaxies are accreted by the cluster as central galaxies. Galaxies that experience pre-processing are those characterised by $t_{\rm lbk,infall} < t_{\rm lbk,infall}^{\rm first}$, and represent the $\sim 57$ per cent of the galaxies accreted into the cluster; although most of these galaxies finally infall to the cluster as satellites, some of them  have been satellites of other haloes in the past and have become centrals later (like BS of other haloes), infalling onto the cluster as central galaxies.

The difference in the hot gas content at the moment of infall of currently passive and star-forming galaxies could indicate the relevance of pre-processing acting at large distances from the cluster centre \citep{mcgee2009,dressler2013,pallero2022}.
Usually, pre-processing is quantified by the fraction of infalling galaxies that are already quenched by the time they arrive to the cluster \citep{Wetzel12,hou2014,haines15, donnari2021a, oman21}. For instance, \citet{haines15} found that the fraction of star-forming galaxies ($f_{\rm SF}$) remains $\sim 20-30$ per cent below the $f_{\rm SF}$ of a coeval field population up to $\sim 3 R_{200}$, and \citet{Wetzel12} found that the radial dependence of $f_{\rm q}$ for all galaxies located around cluster-mass haloes of $M_{\rm 200m}>10^{14} {\rm M_\odot}$ remains higher than average values estimated from field galaxies even at $\sim 10\, R_{\rm 200m}$. 
\citet{jung2018} found evidence of pre-processing by studying the cold gas content of galaxies prior to infall into galaxy clusters: those galaxies that 
belonged to groups as satellite
galaxies before infall (satellite infallers)
are more likely to become depleted compared to those that have always been central galaxies before becoming
satellites of clusters (central infallers). 
In particular, they find that $45$ per cent of satellite infallers become gas poor before cluster entry (specially low-mass galaxies in more massive groups), while only $6$ per cent of central infallers present this behaviour. Although there are numerous contributions in the literature on pre-processing, the direct impact of its
effects beyond \rdos~on properties like SFR, colour or gas content has been marginally explored.

Here, we quantify the impact of RPS during the 
PP stage and CL stage by estimating the fraction of hot gas removed by RPS
during each stage, $f_{\rm RPS}$, as in \citet{Cora_2018}.
This is given by
$f_{\rm RPS} = M_{\rm hot}^{\rm strip,tot}/(M_{\rm hot}^{\rm strip,tot} +M_{\rm hot}^{\rm end})$,
where $M_{\rm hot}^{\rm strip,tot}$ is the total mass of hot gas stripped during a given stage, and
$M_{\rm hot}^{\rm end}$ is the hot gas available at the end of that stage\footnote{
RPS is not the only physical process that affects the hot gas halo.
It is also reduced by gas cooling and increased by cold gas reheated by SN feedback. 
}.
The former is estimated for galaxies of a given mass by accumulating
the amount of hot gas removed by RPS at each snapshot of the simulation
comprised within
each stage, $M_{\rm hot}^{\rm strip,snap}$, that is, $M_{\rm hot}^{\rm strip,tot} = \sum_{\rm snap_{\it i}}^{\rm snap_{\it f}} (M_{\rm hot}^{\rm strip,snap})$, where ${\rm snap}_i$ and ${\rm snap}_f$ are the closest snapshots to the beginning and end of each stage for each galaxy.

We separate the populations of AIN, BS and RIN in galaxies
that experience pre-processing from those that do not, and estimate the fraction $f_{\rm RPS}$ 
corresponding to each subsample in both the PP stage and the CL stage. We present these results in Fig.~\ref{fig:gasstrip}.
We find that pre-processed AIN lose small amounts of gas during the PP stage ($\sim 25$ per cent for $M_\star < 10^{11} {\rm M_\odot}$, solid dark line), consistent with the short timescale of this stage for AIN (left panel). On the contrary, both pre-processed and not pre-processed AIN lose (almost) completely their hot gas during the CL stage
(orange dashed and red dashed-dotted lines, respectively). 
The BS population that experience pre-processing lose $\sim 25$ per cent of their hot gas during the PP stage, and once they enter the cluster they lose $\sim 50$ per cent of their hot gas, regardless of their stellar mass; the BS that are not pre-processed present a behaviour similar to the pre-processed ones during the CL stage (middle panel). This is consistent with the fact that, statistically, BS at $z=0$ have moderate amounts of hot gas 
(see Fig.~\ref{fig:phase-space-hotgas-ratio}).
In the case of RIN (right panel), the duration of the PP stage is expected to be longer than the CL stage, and we find that the accumulated action of RP in smaller haloes results in higher amounts of hot gas stripped during the PP stage ($f_{\rm RPS}\sim 0.5-0.25$, being larger for lower stellar mass) than in the CL stage ($f_{\rm RPS}\sim 0.15$). 
We remark that the values of $f_{\rm RPS}$ for galaxies with $M_\star > 10^{11} {\rm M_\odot}$ during the PP stage are likely to be misleading, as we expect few galaxies as massive as these to be satellites at early stages of the Universe.

It is interesting to explore how RPS during different stages of the evolution affects the final SF
activity of galaxies.
For this, we compute the fraction $f_{\rm RPS}$ for AIN,  
BS and RIN
for the different stages aforementioned, but separating
them in star-forming
and passive galaxies at $z=0$. In the left panel of Fig.~\ref{fig:gasstrip2}, we can appreciate that the PP stage is irrelevant in the final SF activity of AIN, as both star-forming and quenched AIN at $z=0$ lose a small amount of hot gas during this stage (solid dark and dashed orange lines, respectively). On the contrary, we confirm that the RPS of the hot gas during the CL stage is fundamental in quenching AIN, as $f_{\rm RPS}\sim 1$ for quenched galaxies with $M_\star < 10^{10.5} {\rm M_\odot}$ (red dashed dotted line). The fraction of star-forming
AIN at $z=0$ is less than $\sim 20$ per cent and, as they lose $\sim 75$ per cent of their hot gas during the CL stage (dotted greenish line), it is likely that the residual amount of hot gas is still cooling and feeding the cold gas reservoir.

For the BS population (middle panel of Fig.~\ref{fig:gasstrip2}), we find that the currently star-forming BS lose less than $\sim 10$ per cent of hot gas during the PP stage (solid dark line)
and $\sim 40$ per cent during the CL stage (dotted greenish line), regardless of their stellar mass. 
The latter fraction is an intermediate value between those corresponding to AIN and RIN, consistent with the intermediate values of the lookback infall time of BS (Fig.~\ref{fig:histo-time-inside}). 
It is worth mentioning
that $\sim 70$ per cent of currently passive BS experienced pre-processing effects, while $\sim 65$ per cent of currently star-forming BS did not experience pre-processing.
As expected, the moderate action of RP on both the PP and CL stages is also reflected in the population of quenched BS at $z=0$, with $f_{\rm RPS}\sim 0.35-0.5$ (dashed orange line) and $f_{\rm RPS}\sim 0.5-0.7$ (red dashed dotted line), respectively.

In the case of RIN (right panel of Fig.~\ref{fig:gasstrip2}), the impact of RP on the star-forming
population at $z=0$ is weak ($f_{\rm RPS} \lesssim 0.2$) during both PP and CL stages (dark solid and greenish dotted lines, respectively). In fact, $\sim 60$ per cent of currently star-forming RIN have not experienced pre-processing effects at all.
Moreover, the CL stage is almost irrelevant to determine the quenched population of RIN at $z=0$, as $f_{\rm RPS}\lesssim 0.15$ during this stage. 
However, RPS during the PP stage is the responsible for the SF quenching of the low-mass RIN prior to infall, depriving them from their hot gas; likely, AGN feedback is the main cause of SF quenching in high-mass passive
RIN prior to infall. These processes are responsible for $\sim 15$ per cent of passive galaxies inside the clusters, which correspond to the passive RIN.

This simple analysis of the action of RPS on 
different galaxy populations during the PP and CL stages
allows us to connect naturally the properties of galaxies with the main physical processes that affect them through different stages of their evolution, and enable us to explain the SF activity of galaxies located in and around galaxy clusters at the present time. 
Considering the evolution of the fraction of quenched galaxies presented in Fig.~\ref{fig:fq-mstar_AIN-RIN-BS},
we see that, for a given stellar mass, the 
fraction of quenched galaxies
at $z=0$ increases progressively for RIN, BS and AIN (the former might belong to either of the two other populations in the future), but their situation at infall is pretty similar for all of them, with RIN having slightly larger quenched fractions. 
Without the 
analysis described in this section,
one can simply infer from 
the results on the fraction of quenched galaxies
that the role of environmental effects within the clusters is progressively more relevant for RIN, BS and AIN, which is true. However, the important role of the effects of the environment while galaxies orbit within smaller haloes prior to falling into a cluster is not completely captured by the quenched fractions at infall. The analysis presented in Fig.~\ref{fig:gasstrip2} highlights the importance of pre-processing in establishing the galaxy conditions for further action of environmental effects (RPS in this study) inside the cluster. This is particularly relevant to explain the present fractions of quenched RIN and BS.

\section{Discussion}
\label{sec:discussion}

The current version of the semi-analytic model \sag, in combination with the DM-only simulation \textsc{mdpl2}, produces a reliable population of galaxies, where the dependence of $f_{\rm q}$ on stellar mass and halo mass, the gas fractions and cosmic SFR history are in fairly good agreement with observations \citep[see][for details]{Cora_2018,cora2019}.
In particular, the combination of \sag~ with \textsc{The Three Hundred} clusters produces a 
qualitatively good
agreement with observations of the dependence of $f_{\rm q}$ with cluster-centric distance (Fig.~\ref{fig:fq-mstar-r-All-Wetzel14}), a coherent correlation between phase-space position, lookback time since infall and hot gas content (Fig.~\ref{fig:phase-space-hotgas-ratio}), and a consistent evolution of $f_{\rm q}$ and hot and cold gas fractions (Figs.~\ref{fig:fhot-mstar_AIN-RIN-BS} and~\ref{fig:fcold-mstar-AIN-RIN-BS}). 

The current work explores how RPS impacts different populations of galaxies that are in and around massive galaxy clusters, and our results are in broad agreement with \citet{Wetzel_2014}, \citet{jung2018}, \citet{rhee2020} and \citet{oman21}.
In particular, \citet{Wetzel_2014} find that SF quenching occurring in ejected satellites can account for the enhancement of the 
fraction of passive
central galaxies in the outskirts of clusters, leaving little to no room for additional environmental effects outside the virial radius of the clusters. We find that BS require not only the action of environmental processes that occur inside the cluster, but also the hot gas removal that take place during the pre-processing stage in order to be quenched at $z=0$. However, the fraction of quenched galaxies with ${\rm log} (M_\star [{\rm M_\odot}]) \in [9.7,10.5]$ at cluster-centric distances of $r \gtrsim 0.9 R_{200}$ results underestimated with respect to \citet{Wetzel_2014}. 
Moreover, the hot gas fractions of RIN inside \rdos~and BS outside the cluster up to $3$\rdos~ (Figs.~\ref{fig:phase-space-hotgas-ratio} and \ref{fig:fhot-mstar_AIN-RIN-BS}) contrast with the findings of \citet{arthur19} and \citet{mostoghiu21}, where infalling galaxies lose their gas by the time they reach $1.5$\rdos~during their first infall. 
On the one hand, this might
indicate that our RPS model 
does not capture the right efficiency of RP in removing
gas from satellites while they orbit inside massive clusters 
(by uncertainties in either the RP fit employed or the estimation of galaxy properties by \sag), which would result in a 
decrement
of the SF activity in somewhat longer
timescales.
On the other hand, it might be necessary that our RPS scheme includes an explicit dependence on the
environmental local density where the galaxy is located
rather than on a property that relies on the halo finder algorithm, namely, whether a galaxy is satellite or central. 
Several observational studies have shown that environmental effects might act beyond the halo boundary \citep{Wetzel12, lu2012, behroozi2014, arthur19}. Recently, \citet{Ayromlou_2021b} developed a novel gas-stripping method in which central galaxies located in the vicinity of massive haloes ($M_{200}>10^{12} \, {\rm M_\odot}$) can experience RPS. In their model, RPS depends on the properties of the local environment of galaxies \citep{ayromlou2019}, and they show that the dependence of $f_{\rm q}$ on cluster-centric distance, stellar mass and halo mass are in reasonable agreement with observations.
In addition, galaxies can fall into clusters directly from the field or through filaments of the cosmic web. It has been established in \textsc{The Three Hundred} sample that $\sim 20$ per cent of galaxies with $M_\star>10^{9.3}\, {\rm M_\odot}$ around clusters are in filaments \citep{kuchner2020}, and that $\sim 30-60$ per cent of galaxies are environmentally affected by cosmic filaments prior to the first infall onto the cluster \citep{kuchner2022}. 
Although the \sag~model takes into account group-preprocessing, as already mentioned, it does not include the RP effect produced by the gas distributed along filaments that could affect central galaxies located in them.
This could be relevant in the difference between the modeled and the observed $f_{\rm q}$ in the outskirt of the clusters (Fig.~\ref{fig:fq-mstar-r-All-Wetzel14}). 
However, modifying our RPS model and/or the including environmental effects produced by filaments is a task beyond the scope of this work.

We highlight that 
the definition of the halo boundary is key in
the modelling of satellite gas stripping (and, more generally, for any host-subhalo relation), and for the identification
of the BS population. For instance, \citet{Diemer_2021} 
analyse the subhalo fraction and BS 
fraction when different definitions for the halo boundary are adopted: $R_{\rm 200m}$, $ R_{\rm 200c}$, $ R_{\rm vir}$ and splashback radius (which is the distance that contains up to, e.g., $75$ or $90$ per cent of the apocentre of objects that orbit the central halo). The splashback radius has been proposed as a physically motivated definition of the halo boundary \citep{diemer2014,more2015} and can be inferred observationally as it is associated to a sharp drop in the galaxy density field and weak lensing signal around clusters \citep{more2016,mansfield2017,chang2018}. \citet{Diemer_2021} demonstrate that, if the splashback radius is used, the majority of BS would be classified as satellites. 
This somewhat introduces a new interpretation
in which clusters and their environmental effects result naturally extended. 
However, given the characteristics of the \sag~model, we separate the evolution of BS into the stages mentioned in Sec.~\ref{sec:transformation}, and focus on the impact of the pericentric passage. 
The identification of PP and CL stages adopted to evaluate the relative impact of pre-processing and cluster environmental effects are affected by the definition of the cluster boundary, in our case given by \rdos.
About $\sim 20$ per cent of galaxies outside \rdos~are orphan satellites bound to the cluster (see Fig.~\ref{fig:histo-r-Type1Type2} and corresponding analysis). 
Since they are considered satellites, they are affected by RP exerted by the cluster according to the implementation of this process in \sag, despite the fact that they are beyond \rdos. Hence, although the choice of the cluster boundary does not affect RPS of this particular galaxy population, it does affect the classification of them in terms of cluster galaxies or BS.

The pericentric passage is a crucial event in 
the evolution of galaxy properties.
We determine that the majority of star-forming AIN at infall quenches after one pericentric passage, regardless of their stellar mass. AIN quenches between $\sim 3\,{\rm Gyr}$ (low-mass galaxies) and $\sim 1\,{\rm Gyr}$ (high-mass galaxies) after the first pericentric passage, although with a high scatter in the distribution, which allows a non-negligible fraction of AIN to reach a second pericentric passage before SF quenching (see Fig.~\ref{fig:tdelay}). This is in general consistent with \citet{oman21}, although our delay times are systematically lower at all stellar masses. 
We mention that our global quenching timescale for AIN and BS is $\sim 0.5 \, {\rm Gyr}$ higher than their $t_{\rm delay}$ at fixed stellar mass, which is consistent with the findings of \citet{rhee2020}, but in contrast with \citet{Lotz_2019} and \citet{Upadhyay_2021}, who find shorter quenching timescales of only $\sim 1 \, {\rm Gyr}$ after infall and $\lesssim 1\, {\rm Gyr}$ after pericentric passage, respectively, consistent with the truncation of SF around the first pericentre. 
\citet{steinhauser2016} also find short quenching timescales ($\lesssim 1\, {\rm Gyr}$), directly associated to the fast depletion of the hot gas halo (in timescales of $\sim 200\,{\rm Myr}$ for a galaxy of $\sim 10^{10}\, {\rm M_\odot}$ in halos of $\sim 10^{15}\, {\rm M_\odot}$) as a result of RPS. 

\citet{jung2018} use the YZiCS set of hydrodynamical simulations of galaxy clusters to study when and where the depletion of cold gas takes place, and the timescales involved in the process. 
They find that overall $\sim 70$ per cent of galaxies are still gas rich at the cluster entry, and they lose their gas content mainly by the action of RPS in $\lesssim 1\,{\rm Gyr}$ in massive clusters ($M_{200} > 10^{14.5}\, {\rm M_\odot}$). This means that galaxies are depleted of cold gas before reaching their first pericentric passage, which contrasts with our findings. As expected, 
the timescale in which the gas reservoir of a galaxy is depleted depends on both the amount of gas at infall and the shape of the orbit: satellites with less gas at infall and with more radial orbits lose their gas faster. The other $\sim 30$ per cent of galaxies are already gas-poor before cluster infall, specially low-mass satellite galaxies that inhabit massive groups prior to infall. This supports the scenario in which RPS can remove the gas content of galaxies in groups before being accreted
into a cluster, in clear consistency with our findings. 
We also note that in our work, the study of AIN, BS and RIN provides different evolutionary tracks to produce quenched galaxies in and around galaxy clusters. This evolutionary tracks are consistent with the scheme of three channels that produce gas-poor galaxies inside clusters, proposed by \citet{jung2018}.
Although we use a different numerical technique and RPS model, our results on the AIN and BS populations support the picture in which the removal of the hot gas halo is a necessary condition for SF quenching to occur in low-mass galaxies ($M_\star \lesssim 10^{10.5}\, {\rm M_\odot}$), and we probe it over a larger statistical sample of galaxies.

\section{Conclusions}
\label{sec:conclusions}

In this work, we apply the semi-analytic model of galaxy formation and evolution \sag~to the DM-only simulations of galaxy clusters of \textsc{The Three Hundred} project to provide a semi-analytic approach on 
the interpretation of
the SF quenching in galaxies
and its connection to their dynamical evolution in high density environments. 
Our model includes the information obtained from tracking the orbital evolution of unresolved subhaloes that are populated with orphan satellites by \sag. 
We restrict our analysis to the $\sim 102$ most relaxed clusters of \textsc{The Three Hundred} sample, and
we define three populations of galaxies, based on their current location and orbital evolution: ancient infallers (AIN, current satellites that have crossed \rdos~of the cluster more than $2\,{\rm Gyr}$ ago), recent infallers (RIN, current satellites that have crosses \rdos~of the cluster less than $2\,{\rm Gyr}$  ago), and backsplash galaxies (BS, galaxies that have crossed \rdos~at some point of their evolution, and currently lie outside the cluster); we select galaxies with $M_\star \geq 10^9\, {\rm M_\odot}$ in order to avoid resolution effects. 

Our main findings are summarized as follows:

\begin{itemize}

    \item Passive AIN represent $\sim 85$ per cent of the current passive population inside clusters, and present low or null content of hot and cold gas, depending on their stellar mass. The fraction of passive AIN increases strongly between first and second pericentric passages (Fig.~\ref{fig:fq-mstar_AIN-RIN-BS}, left panel), 
    with a simultaneous 
    decrease of their hot gas content (Fig.~\ref{fig:fhot-mstar_AIN-RIN-BS}, left panel), a trend more pronounced for less massive galaxies. The depletion of hot gas is a necessary condition for low-mass AIN to be quenched at $z=0$, and this is a consequence of the action of RPS over long periods of time exerted by the cluster itself; the impact of pre-processing is minor (Fig.~\ref{fig:gasstrip2}, left panel). Moreover, as the hot gas halo is removed, RPS can act over the cold gas phase, which is completely removed by ${z=0}$ (Fig.~\ref{fig:fcold-mstar-AIN-RIN-BS}, left panel).
    
    \item 
    Only $\sim 30$ per cent of RIN are passive at ${z=0}$; passive RIN represent only $\sim 15$ per cent of the current passive population inside the clusters.
    In general, star-forming RIN still preserve a high proportion of hot gas, which correlates with their position in the (real) phase-space diagram.
    Star-forming RIN with high relative velocities and low cluster-centric distances can retain a hot gas halo
    (Fig.~\ref{fig:phase-space-hotgas-ratio}, top right panel),
    even after experiencing a pericentric passage ($\sim 50$ percent of RIN have reached the pericentre of their orbit). The fraction of RIN that experience a pericentric passage is higher than the fraction of quenched RIN, which indicates that a timescale $\lesssim 2$ Gyr is not enough for RPS to efficiently remove gas 
    in environments with cluster-like densities
    and shut down the SF activity. 
    Indeed,
    $f_{\rm q}$ has a negligible evolution between infall and ${z=0}$ (Fig.~\ref{fig:fq-mstar_AIN-RIN-BS}, middle panel). For passive RIN, the amount of hot gas removed by RPS in smaller haloes in a pre-processing stage (before being accreted into the cluster) results higher than inside the cluster itself (Fig.~\ref{fig:gasstrip}, middle panel).
    
    \item Passive BS represent $\sim 65$ per cent of the overall passive population in the outskirts of clusters, between $1-3$ \rdos. Even though the whole population of BS experienced a pericentric passage, 
    around half of it still preserves a moderate proportion of hot gas, which does not correlate strongly with the position on the phase-space diagram. When we restrict to those BS that are quenched at ${z=0}$, we find that the decrease of their hot gas content results more abrupt than for the star-forming population of BS (Fig.~\ref{fig:fhot-mstar_AIN-RIN-BS}, right panel), and the depletion of hot gas is
    a necessary condition for SF quenching to occur. 
    Neither the distance to the cluster centre during the first passage nor the time spent inside the cluster determines the final state of BS at ${z=0}$. Actually, the stage in which they are satellites of other haloes prior to infall onto the cluster (i.e., the pre-processing stage) is critical in the removal of a proportion of their hot gas (Fig.~\ref{fig:gasstrip2}, right panel). Thus, the combination of the action of RPS during the pre-processing stage and the cluster stage sets the conditions for the BS to be depleted of hot gas, and hence, be quenched at $z=0$.

\end{itemize}

The majority of our AIN quenches between first and second pericentric passages, regardless of their stellar mass content, although a relevant proportion of AIN with $M_\star < 10^{10.5}\, {\rm M_\odot}$ quenches between second and third passages. 
The latter situation
could be associated to satellites with early accretion times, when the ICM was not dense enough to efficiently remove gas from them. However, given that our model underestimates $f_{\rm q}$ for galaxies of $M_\star < 10^{10.5}\, {\rm M_\odot}$ located at $r\gtrsim 0.8\,R_{200}$, it might indicate that the RPS model implemented is not being strong enough to deplete low-mass galaxies from their gas content in timescales of $\sim 1.5-2 \,{\rm Gyr}$.
For those AIN that are star-forming at infall and quenched at ${z=0}$, we find that they quench $\sim 3$ ($\sim 1 \, {\rm Gyr}$) after first pericentric passage for galaxies with low (high) stellar mass, although with a high scatter in the distribution (Fig.~\ref{fig:tdelay}), likely associated to the different orbital trajectories of the sample. 
This delay time results somewhat shorter for low-mass BS,
as $t_{\rm delay}\sim 2\,{\rm Gyr}$, while 
high-mass BS have similar delay times as high-mass AIN ($t_{\rm delay}\sim 1\,{\rm Gyr}$), again with a wide scatter.)

Our results contribute to the general picture of galaxy evolution in the context of high density environments, in particular,
the relationship of their
hot gas content and SF activity
with their orbital history. 
Many works in the literature (see Sec.~\ref{sec:intro}) support RP as the dominant process responsible for SF quenching, and we arrive to similar conclusions with \sag.
It is worth noting that the implementation of RP in \sag~does not consider it as a mechanism for triggering SF as a result of gas compression \citep{roberts2020}, which could mitigate its SF quenching effects. Having these caveats in mind, the interpretation of our model results also emphasizes the need to take into account the pre-processing stage prior to accretion into the cluster, and not only the cluster stage itself, in order to interpret the evolution of 
galaxy properties within clusters and in their outskirts.

As we mention in Sec.~\ref{subsec:gascontent}, the correlation of the hot gas content with the position in the phase-space diagram of galaxies from \sag~(Fig.~\ref{fig:phase-space-hotgas-ratio}) and \textsc{GadgetX} \citep[fig. 6 of][]{mostoghiu21} is very different. 
We can infer that the physical mechanisms that remove gas from galaxies in \textsc{GadgetX} seems to be rather extreme, as infalling galaxies that cross $1.5 R_{200}$ become completely depleted of gas by $z=0$, regardless
of their time-since-infall.
This contrasts with several observational results that show the
milder effects of environmental processes (as we mention in Sec.~\ref{sec:intro}), and 
with the results presented in this work.
We refer the reader to \citet{Cui_2018}, where a first approach to a comparison between different SAMs (\sag, \textsc{sage} and \textsc{galacticus}) and hydro runs (\textsc{GadgetX} and \textsc{GadgetMUSIC}) is shown.
A detailed comparison
between results of \sag~and hydrodynamical simulations such as \textsc{GadgetX} and \textsc{gizmo-simba} \citep{cui2022}, analysing not only the gas content, but also the SFR that is determined by the circulation of baryons, would help to elucidate the strengths and weaknesses of both approaches in understanding the physical processes that trigger SF quenching in galaxy clusters and their outskirts. Such a comparison is planned for a future work.


\section*{Acknowledgements}

We thank the referee for helpful comments and suggestions.
This work has been made possible thanks to `The Three Hundred' collaboration. The project received financial support from the European Union's Horizon 2020 Research and Innovation Programme under the Marie Sklodowska-Curie grant agreement No 734374. Project acronym: LACEGAL.
TH acknowledge {\it Consejo Nacional de Investigaciones
Cient\'{\i}ficas y T\'ecnicas} (CONICET), Argentina, for their supporting fellowships.
SAC acknowledges funding from  CONICET (PIP-2876), {\it 
Agencia Nacional de Promoci\'on de la Investigaci\'on, el Desarrollo Tecnol\'ogico y la Innovaci\'on} (Agencia I+D+i, PICT-2018-3743), and {\it Universidad Nacional de La Plata} (G11-150), Argentina.
CVM acknowledges support from ANID/FONDECYT through grant 3200918, and he also acknowledges support from the Max Planck Society through
a Partner Group grant.
Most of the simulations used in this work  have been performed in the MareNostrum Supercomputer at the Barcelona Supercomputing Center, thanks to CPU time granted by the Red Espa\~nola de Supercomputaci\'on.
WC, AK  and GY would like to thank Ministerio de  Ciencia e Innovación (Spain) for partial financial support under research grant PID2021-122603NB-C21.
AK also thanks Breeders for cannonball.
The programs and figures had been developed under \textsc{Python} language \citep{Python3}, using specially \textsc{NumPy} \citep{Van2011numpy} and \textsc{Matplotlib} \citep{Hunter2007}.
%

\section*{Data Availability}

This manuscript was developed using data from \textsc{The Three Hundred} galaxy clusters sample. The data is available on request following the guideline of \textsc{The Three Hundred} collaboration (\url{https://www.the300-project.org}).
The data underlying this article will be shared on reasonable request to the corresponding author.



\bibliographystyle{mnras}
\bibliography{references}





\bsp	
\label{lastpage}
\end{document}